\documentclass[useAMS,usenatbib]{mnras}

\usepackage{natbib}
\usepackage{graphicx}
\usepackage{rotating}
\usepackage{times}
\usepackage{amsmath}
\usepackage{color}

\definecolor{darkgreen}{rgb}{0.0,0.5,0.0}
\definecolor{orange}{rgb}{1,0.5,0}
\definecolor{darkred}{rgb}{0.75,0.,0.2}
\definecolor{magenta}{rgb}{0.8,0,0.8}
\definecolor{purple}{rgb}{0.5,0,0.5}
\definecolor{gray}{rgb}{0.5,0.6,0.7}
\definecolor{darkblue}{rgb}{0.1,0.1,0.75}

\title[Simulating Galactic Dwarfs Formation]{The Accretion History of the Milky Way: III. Hydrodynamical Simulations of Galactic Dwarf Galaxies at First Infall} 

\author[Jianling Wang et al.]{
Jianling Wang $^{1,2}$\thanks{E-mail:wjianl@bao.ac.cn}, Francois Hammer$^{1}$\thanks{E-mail:francois.hammer@obspm.fr},
Yanbin Yang$^{1}$, Marcel S. Pawlowski$^{3}$, 
 \newauthor
 Gary A. Mamon$^{4}$, Haifeng Wang$^{5}$  
\\
$^{1}$GEPI, Observatoire de Paris, Paris Sciences et Lettres, CNRS, Place Jules Janssen 92195, Meudon, France.\\
$^{2}$CAS Key Laboratory of Optical Astronomy, National Astronomical Observatories, Beijing 100101, China\\
$^{3}$Leibniz-Institut fuer Astrophysik Potsdam (AIP), An der Sternwarte 16, D-14482 Potsdam Germany\\
$^{4}$Institut d'Astrophysique de Paris (UMR7095: CNRS \& Sorbonne Universit\'e), 98 bis Bd Arago, 75014, Paris, France\\
$^{5}$CREF, Centro Ricerche Enrico Fermi, Via Panisperna 89A, I-00184, Roma, Italy
}

\begin{document} 

\date{Received ; accepted}

\maketitle

\begin{abstract}
Most Milky Way dwarf galaxies are much less bound to their host than are relics of Gaia-Sausage-Enceladus and Sgr. These dwarfs are expected to have fallen into the Galactic halo less than 3 Gyr ago, and will therefore have undergone no more than one full orbit. Here, we have performed hydrodynamical simulations of this process, assuming that their progenitors are gas-rich, rotation-supported dwarfs.  We follow their transformation through interactions with the hot corona and gravitational field of the Galaxy.
Our dedicated simulations reproduce the structural properties of three dwarf galaxies: Sculptor, Antlia II and, with somewhat a lower accuracy, Crater II. This includes reproducing their large velocity dispersions, which are caused by ram-pressure stripping and Galactic tidal shocks. Differences between dwarfs can be interpreted as due to different orbital paths, as well as to different initial conditions for their progenitor gas and stellar contents. However, we failed to suppress in a single orbit  the rotational support of our Sculptor analog if it is fully dark-matter dominated.
In addition, we have found that classical dwarf galaxies like Sculptor may have stellar cores sufficiently dense to survive the pericenter passage through adiabatic contraction. On the contrary, our Antlia II and Crater II analogs are tidally stripped, explaining their large sizes, extremely low surface brightnesses, and velocity dispersion. This modeling explains differences between dwarf galaxies by reproducing them as being at different stages of out-of-equilibrium stellar systems.
\end{abstract}


\section{Introduction}

The {\it Gaia} mission's measurements of the bulk proper motions of  Milky Way (MW) dwarf galaxies have opened a new avenue to understand these low mass stellar systems. \citet[hereafter Paper~I]{Hammer2023} established an empirical relation between the infall lookback time and the logarithm of the orbital energy by comparing the total orbital energy of dwarfs to that of events with well-known infall time (Gaia-Sausage-Enceladus, 8-10 Gyr, Sgr, 4-6 Gyr ago). This empirical relation is in excellent agreement with theoretical predictions from cosmological simulations \citep{Rocha2012}, and it suggests that  most dwarfs arrived recently ($<$ 3 Gyr ago). 
The late infall of dwarfs requires reconsidering\footnote{Otherwise, if dwarfs were accreted a long time ago, most of their progenitor properties would have been diluted during the numerous subsequent orbits that would have let enough time to reach equilibrium.}  the properties of their progenitors outside the MW halo, which are likely gas-rich dwarf galaxies. This is because most dwarfs outside MW or M31 halos are dwarf irregulars (dIrrs) as shown by \citet{Putman2021}, 
and their passage into the MW halo gas may have fully transformed them \citep{Mayer2006,Yang2014}. This is expected because (1) of the considerable loss of mass associated to the gas removal through ram pressure from the MW halo gas \citep{Yang2014}, (2) the strong turbulence caused by this gas extraction affecting both gas and stellar kinematics, and (3) of Galactic tidal shocks \citep[hereafter Paper~II]{Hammer2024} mostly occurring at pericenter, a location close to which many dwarf galaxies are found \citep{Fritz2018,Li2021}.  \\

The presence of a hot corona surrounding the MW is necessary to explain the dual HI filaments of the Magellanic Stream \citep[e.g.,][]{Hammer2015,Wang2019}, while the Leading Arm has been suggested to be generated by satellites moving ahead of the LMC \citep{Tepper-garcia2019}. This hot gas exerts ram-pressure on the dwarf progenitors and strip their gas. Other evidences for the presence of MW hot gas are coming from (1)  the dichotomy of gas content in gas-poor (rich) dwarfs inside (outside) the MW or M31 halo \citep{Grcevich2009,Putman2021}, and (2) the gaseous disk of the Large Magellanic Cloud (LMC) that has shrunk enough to be smaller that the stellar counterpart suggesting very strong ram-pressure effects \citep{Nidever2014,Salem2015}, which make it rather unique among other dIrrs. Detection of the high velocity cloud dissociations at distances larger than 50 kpc by \cite{Kalberla2006} implies densities of few $10^{-4}$ $\rm cm^{-3}$, which are consistent with all other estimates \citep{Salem2015,Hammer2015,Wang2019,Putman2021}.


The above paragraphs suggest that most dwarf properties result from a temporal sequence, beginning with gas stripping due to MW halo gas ram-pressure, expansion of their stars due to the subsequent lack of gravity, and then a significant impact of MW tidal shocks exerted mostly on the leaving stars. The impact of such an out-of-equilibrium process has been theoretically described by Hammer et al. (2023b hereafter Paper~II). The latter shows that dwarf velocity dispersions are likely increased by gas turbulence, expansion of the stellar distribution and tidal shocks. It comes out that most MW dwarfs as they are observed today are not in equilibrium, and that the self-equilibrium conditions assumed by \citet{Walker2009} and \citet{Wolf2010} cannot be applied to them. 

The goal of this paper is to verify whether one can reproduce through hydrodynamical simulations the properties of MW dwarf galaxies after assuming they are at their first infall in the MW halo. Section~\ref{sec:initial} describes the hydrodynamical model and the initial conditions for the MW and dwarf progenitors including their gas content (Section~\ref{sec:gas}).  Section~\ref{sec:DM} shows that we failed to reproduce a dispersion-supported Sculptor analog when its progenitor is dark-matter dominated. 
Section~\ref{sec:simus} shows the simulations of Sculptor, Antlia II and Crater II analogs, and compares the results to observations.  The differences in size and velocity dispersion between these three dwarfs can be simply explained by their different orbital paths. Section~\ref{sec:simus} also investigates the numerical convergence issues, especially those related to the gas removal conditions. Section~\ref{sec:discussion} discusses the different mechanisms that explains the different properties of the simulated dwarf galaxies together with predictions for future observations, while Section~\ref{sec:conclusion} summarizes the main results of the study.




\section{Hydrodynamical model and Initial conditions for the Milky Way and dwarf progenitors}
\label{sec:initial}
\subsection{Hydrodynamical model} 

Numerical simulations have been carried out with GIZMO \citep{Hopkins2015}, which is based on a new Lagrangian method for hydrodynamics, and has simultaneously properties of both smoothed particle hydrodynamics (SPH) and grid-based/adaptive mesh refinement (AMR) methods.  It has considerable advantages when compared to SPH: proper convergence, good capturing of fluid-mixing instabilities, dramatically reduced numerical viscosity, and sharp shock capturing.  These features make GIZMO much more advanced to capture hydrodynamics than GADGET-2 \citep{Springel2005}, which is unable to
properly account for, e.g., Kelvin-Helmhotz instabilities.

\subsection{Initial conditions for the Milky Way and its halo gas}
\label{sec:MWmass}
The MW model is set up from that used in \citet{Wang2019}, which reproduces well the Magellanic System including the Magellanic Stream, Clouds, and Bridge. Many predictions from this model have been confirmed by recent observations \citep{Wang2022}.  \citet{Wang2019} used two similar MW models, which includes a stellar disc, gas disc, hot gas corona,
and dark matter halo.  The hot corona density profiles are slightly different in these two models, but both of them can reproduce well the Magellanic System \citep{Wang2019}. In this work, we 
adopt model MW1 of \citet{Wang2019}, which has a total dynamical mass of 6.9 $\times10^{11}M_{\odot}$ (see \citealt{Wang2019} for details). The total number of particles in the MW model is around 30
million, for which masses of dark matter and corona gas particles are 10$^5$ M$_{\odot}$.


\begin{figure}
\centering
\includegraphics[width=9cm]{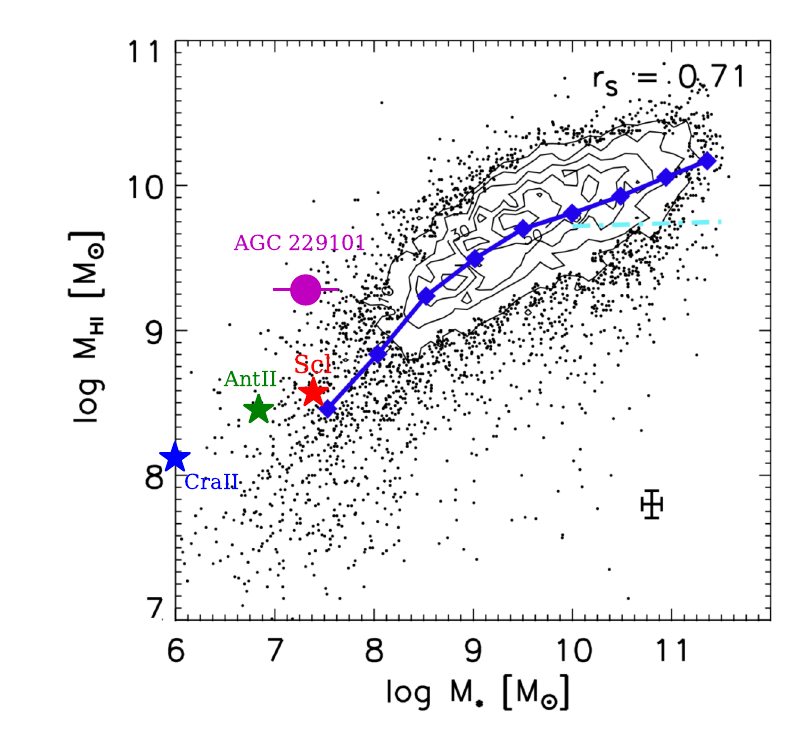}
\caption{Stellar mass versus H{\sc i} mass distribution for ALFALFA sample (black dots and contours) from
\citet{Huang2012}, to which we have added the progenitors of Crater II (blue star),
Antlia II (green star), and Sculptor (red star).  Blue diamonds and solid lines
indicate the mean values in each stellar mass bin. The cyan dash-dotted line is the average
value of the GASS galaxies with M$_\star \ge 10^{10}$ M$_{\odot}$ from \citet{Catinella2010}.
The Spearman rank correlation coefficients ($r_S$, \citealt{Huang2012}) is labelled on the top right.
The almost dark galaxy AGC 229101 from \citet{Leisman2021} is indicated by a magenta circle.}
\label{HImass}
\end{figure}


\begin{table*}

\caption{Initial condition parameters and particle numbers for the progenitors of three Galactic dwarfs.}

\begin{tabular}{|l|c|c|c|c|}

\hline

Dwarf Model                                             & DM-dominated Sculptor  &  Sculptor      & Antlia II        &  Crater II                    \\

\hline

\hline

Gas disc mass ($10^8$M$_{\odot}$)                       & 0.3  &    6.0         &      4.5         &  2.0                     \\

\hline

Stellar component  mass ($10^6$M$_{\odot}$)             & 5  &    20.         &      7.0         &  1.0                     \\

\hline

Stellar core       mass ($10^6$M$_{\odot}$)             & $-$  &    5.          &   $-$            &      $-$                   \\

\hline

Scale radius of gas Disk (kpc)                          & 0.56  &    1.5         &  2.0             &      2.0                      \\

\hline

Projected effective radius of stellar component (kpc)   & 0.279 &    0.5         &      0.5         &  0.3             \\

\hline

Sersic index ($n$) of stellar component                 &  1 &    0.5         &      1.0         &  0.77                   \\

\hline

Axis ratio of stellar component                         & 0.6  &    0.6         &  1.0             &      1.0           \\

\hline

Number of Stellar Particles                             &  3000 &   12000        &    7500          & 6000                   \\

\hline

Number of Gas Particles                                 & 3000  &    60000       &      68000       & 60000               \\

\hline

\end{tabular}

\label{tab:IC}

\end{table*}

\subsection{Initial conditions for dwarf progenitors}
\subsubsection{Preliminary investigations and stability of initial conditions}
\label{sec:initial_dwarf}
The progenitors of MW dwarf galaxies are assumed to be gas-rich dwarfs, and they include a rotationally-dominated, thick-disk stellar component and a disk gas component. In this paper, we initially focus on reproducing the properties of the Sculptor dwarf spheroidal galaxy (dSph) that is ranked the third in mass among classical dSphs, which also helps to keep a reasonable mass ratio between MW halo gas particles and dwarf gas particles ($m$=10$^4$ M$_{\odot}$). Sculptor is an interesting dSph to study because it has been often considered as an archetype for an ancient satellite of the MW, due to its old stellar population \citep{deBoer2012}. During our preliminary tests, we have found that a single stellar component is very fragile to the gas loss and to Galactic tidal shocks. It results in a very low surface brightness remnant, reminiscent of Antlia II and Crater II. Therefore, we have added a cored stellar component with an exponential profile for the Sculptor progenitor, which has a stellar mass of $5\times10^6$ M$_{\odot}$ together with a half mass radius of 0.15 kpc (see Table \ref{tab:IC}).  This is consistent with the WLM dwarf stellar-density profile, whose thick disk requires two Sersic profiles to be reproduced \citep{Higgs2021}. \\
Here, we have not considered reproducing dwarfs less massive in stars, including 10 times smaller dSphs such as UMi, Draco or Carina, or even ultra-faint dwarfs, because it would have led to uncomfortably high mass ratios of MW halo gas to dwarf gas particles (see Section~\ref{sec:convergence}).
Initial conditions have been created with a Schwarzschild orbit superposition method \citep{Vasiliev2013, Vasiliev2019}. To test the stability of the initial conditions, we have let the dwarf progenitors evolve in isolation for 2 Gyr. Figure~\ref{ICs} of Appendix~\ref{sec:ICs} shows the time evolution of the projected stellar surface mass density and of the projected velocity dispersion profiles. Their cores are mostly preserved during their evolution in isolation, although their outskirts are affected especially by feedback processes, which does not affect our simulation results when launching them to interact with the MW halo and its corona. 


\subsubsection{Stellar and gas mass of the progenitors}
\label{sec:gas}
\citet{Huang2012} compared the H{\sc i} mass versus stellar mass relation for
local galaxies (see Figure~\ref{HImass}). They found that many field dwarfs are gas-dominated, with stellar mass ranging from $10^6$
M$_\odot$ to $2\times10^8$ M$_{\odot}$, and H{\sc i} gas mass ranging from 10$^8$ M$_{\odot}$ to 10$^{10}$ M$_{\odot}$. \\

We have used the average values of local galaxies (see Figure~\ref{HImass}) to capture the properties of the Sculptor progenitor. However, we notice that gas-rich dwarfs lost a significant part of their stellar content during their infall and gas-stripping. This is because after gas exhaustion, the stellar content expands due to the related lack of gravity and it becomes easily affected by MW tides (see description in Paper~II). We have then chosen the Sculptor progenitor with a stellar mass of $2.5\times10^7$ M$_{\odot}$\footnote{However, during our test of dark matter properties (see Section~\ref{sec:DM}, we have also used stellar mass of $5\times10^6$ M$_{\odot}$, i.e., that of the Sculptor dSph.}, leading to an H{\sc i} mass $M_{\rm HI}$= $6\times10^8$ M$_{\odot}$ (see Table \ref{tab:IC}), if it follows the $M_{\rm HI}$-$M_{\rm stars}$ relation of Figure~\ref{HImass}. The progenitor of Sculptor is then assumed to have a 96\% gas fraction (see Table \ref{tab:IC}). This is comparable to that of the NGC 3109 dwarf. The latter has a stellar mass of $7.6\times10^7$ M$_{\odot}$, and a H{\sc i} mass of
$4.5\times10^8$ M$_{\odot}$ \citep{McConnachie2012}, which corresponds to a 90\% gas fraction when accounting for Helium. \\

Given our promising preliminary tests, we have also tried to reproduce Antlia II and Crater II with a single stellar component. They possess stellar masses approximately 10 times smaller than that of Sculptor. We have considered the possibility of an extremely large gas fraction for the progenitors of Antlia II and Crater II, which would strongly impact their stellar content after the gas has been removed by ram pressure. We have then assumed 99.5\% and 98.5\% for the gas fraction of Antlia II and Crater II progenitors, respectively (see Table~\ref{tab:IC}).  Such very high gas fraction has been found by \citet{Leisman2021} when analyzing the unusual HI-dominated galaxy, AGC229101, which has been observed with the Arecibo Legacy Fast ALFA (ALFALFA). Its H{\sc i} mass is 10$^{9.31\pm0.05}$ M$_{\odot}$ with a stellar mass of 10$^{7.32\pm0.33}$ M$_{\odot}$, which leads to a neutral gas fraction of 98\%, a value quite similar to that of Coma P  \citep{Leisman2021}. \\

Figure~\ref{HImass}
compares the distribution of stellar mass and H{\sc i} mass of our dwarf progenitors with that of observed local galaxies. It indicates that the Sculptor progenitor lies well in the expected region for a galaxy with that stellar mass, while Antlia II and especially Crater II progenitors, are on the high-gas fraction side of the distribution, together with AGC229101. Table~\ref{tab:IC} lists the progenitor properties of Antlia II and Crater II. Progenitors of Antlia II and Crater II are more exceptional than that of Sculptor, consistent with the fact that the latter is more representative of MW dwarf galaxies than the former two dwarfs \citep{Ji2021}.

\subsubsection{The role of dark matter and constraints on its content in the Sculptor progenitor}

\label{sec:DM}
\begin{figure*}
\centering
\includegraphics[width=17cm]{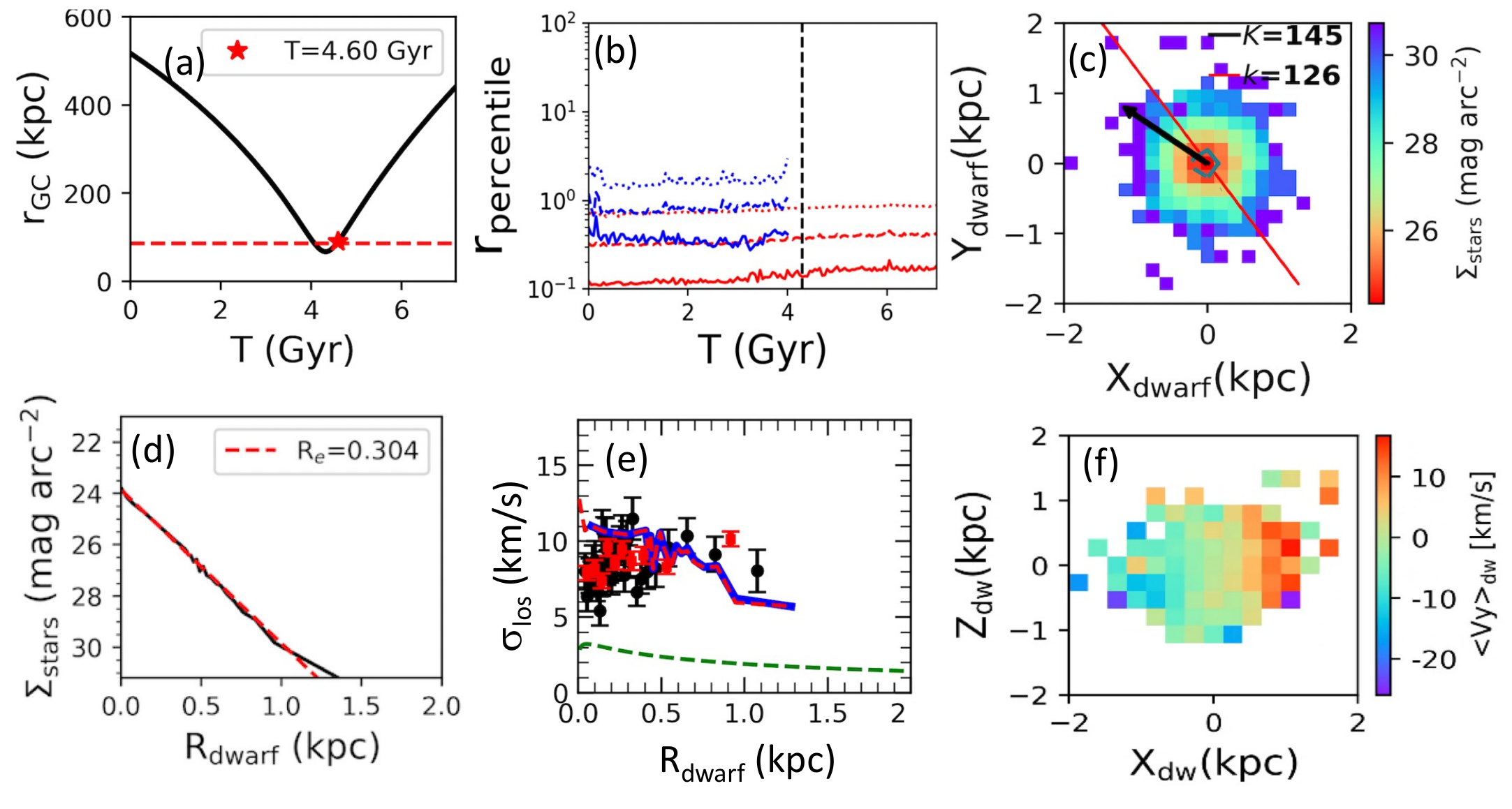}
\caption{ Simulation model of a dark matter-dominated, gas-rich dwarf infall into the Galactic corona, with the gas being removed close to pericenter. This model has the goal to reproduce the Sculptor dwarf after assuming that its progenitor is embedded into a massive dark matter halo with a mass calculated from \citep{Wolf2010}. 
Panel (a): The orbital history of the simulated dwarf (black line), for which the red star indicates the present-day dwarf position. At this time, the gas has been stripped by ram-pressure. 
Panel (b): Evolution of the radii containing 10\% (solid line), 50\% (dashed line) and 90\% (dotted-line) of the total mass, respectively. Red (blue) lines indicate the stellar (gas) component, respectively. 
The vertical line indicates the time at pericenter passage.
Panel (c): Surface brightness profile overlapping contours showing the stellar shape of the dwarf. The red line shows the major axis. The black arrow indicates the proper motion direction, which is 19 degrees offset from the major axis. 
Panel (d): The black line shows the stellar surface brightness profile assuming $M_{\rm stars}/L_{\rm V}=2.5$, while the dashed red line indicates an exponential profile fitting the central region (half-light radius value is indicated on the top-right of the panel). 
Panel (e): Radial velocity dispersion distribution as a function of the radius. The blue line shows the measured values, while the red-dashed line shows results after subtracting the radial velocity gradient as it is done for observed quantities \citep{Walker2009}. The dashed green line shows the contribution to the radial velocity dispersion that is due to the gravity of the single stellar mass.The observed velocity dispersions at different radii are shown by solid black circles for \citet{Walker2009} and with red dots for \citet{Iorio2019} observations, respectively. 
Panel (f): It shows the dwarf as it would be seen edge-on to illustrate its mean velocity along the Y direction, showing a significant rotational gradient that is much larger than that observed in Sculptor.
}
\label{Scl_DM3}
\end{figure*}

MW dSph galaxies have been considered to be dark-matter (DM) dominated, assuming that their large velocity dispersions are due to equilibrium with their total mass \citep{Walker2009,Wolf2010}. This scenario is consistent with a very early infall of their gas-rich and rotationally-supported progenitors, offering a very large elapsed time during which they can be transformed into dispersion-supported dSphs \citep{Mayer2006}. Conversely to that, the large orbital energy of dwarf galaxies when it is compared to that of Sgr or GSE relics, suggests that they come recently, less than 3 Gyr ago, to our neighborhood. In such a case, one has to test whether a Sculptor progenitor that is rotating and DM dominated could be transformed into a dwarf spheroidal within less than one orbit. We have adopted a total mass derived from \citet[see their Equation 2]{Wolf2010} with $\sigma_{\rm los}$= 9.2 km $s^{-1}$, which leads to $2.2\times10^7$ M$_{\odot}$ within a 279 pc half-light radius\footnote{We have adopted a NFW model \citep{Navarro1997} for the dSph, and in our third model shown in Figure \ref{Scl_DM3}, the number and mass of dark matter particles are $10^6$ and 5023 $M_{\odot}$, respectively.}. Extrapolating it to large radii, \citet{Read2019} calculated a halo mass of $5.7\times 10^9$ M$_{\odot}$. We have also adopted the orbital parameters of Sculptor, that are discussed in Section~\ref{sec:orbits}. This has resulted in a DM gravity being sufficiently large to prevent any gas removal during the first passage and up to the present-day Sculptor position. \\
To investigate under which conditions the gas can be removed after a single pericenter passage, we have then considerably diluted the gas in the progenitor by dividing its mass by a factor 10 while keeping the same scaling (see Table~\ref{tab:IC}). However, this second model was not predictive of the observed stellar mass, because the DM component is sufficiently massive to shield the stellar component that is mostly preserved. This has prompted us to consider a third DM-dominated progenitor, with an initial stellar mass of $5\times10^6$ M$_{\odot}$ (i.e., that of the Sculptor dwarf) that follows an exponential profile with a half mass radius of 0.279 kpc. The gas fraction (85\%) has been adopted following the observational constraint from Figure \ref{HImass}. \\
The gas component has been assumed with a scalelength 3 times larger than that of the stellar component. In the third model, the gas is removed near pericenter (see panels (a) and (b) of Figure \ref{Scl_DM3}), and becomes rapidly ionized since it was already heated in the dwarf due to the strong DM gravity, explaining why no more neutral gas appears in panel (b) at T$>$ 4 Gyr. 

Panel (b) also shows that the stellar component stays almost untouched during the whole simulation, which is confirmed by panel (d) showing  a surface brightness profile very similar to the initial one (same exponential profile and half-light radius). Panel (e) provides the velocity dispersion radial profile, which is likely due to the internal equilibrium with the total mass as it has been assumed from Equation 2 of \citet{Wolf2010}. It implies that most properties of Sculptor are reproduced this way, simply because the stellar content and its velocity dispersion stayed untouched because they were shielded by the dominant DM component.

However, panel (f) indicates a significant rotation pattern (rotational velocity larger than 10 km $s^{-1}$), after placing  the dwarf edge-on, conversely to observations that show Sculptor kinematics fully dominated by velocity dispersion (and a velocity gradient smaller than a few km $s^{-1}$, see Section~\ref{sec:Sculptor}). We have not been able to find a DM-dominated Sculptor progenitor that would be consistent with the present-day Sculptor kinematics, as well as with the presence of stars in the very outskirts of dwarfs. This is because a dominant DM component always shields the stellar component and its kinematics against perturbation due to gas loss and Galactic tides. This is in agreement with simulations made by \citet{Mayer2006} who found that several orbits are necessary to successfully realize a transformation from a rotation-supported to a dispersion-supported stellar system. It does not mean that MW dwarfs do not possess DM, but simply that a recent infall dramatically reduces the parameter space against large amounts of DM. \\
Moreover, we notice that panel (a) of Figure \ref{Scl_DM3} indicates a very large apocenter in excess of 400 kpc, or in other words that the dwarf has not been captured by the MW gravitational potential. Given the total mass of the MW adopted in our modeling (6.8 $\times 10^{11}$$M_{\odot}$, see Section~\ref{sec:MWmass}) conversely to {\it Gaia} EDR3 measurements from \citet{Li2021} that indicates apocenter values between 133 to 209 kpc. In fact, in a DM-dominated dwarf, gas-removal becomes an almost negligible event, which prevents any slowdown associated to the gas removal. In addition, if dwarfs are DM-dominated dwarfs, the anti-correlation between DM density and pericenter may be difficult to understand, while it is predicted if they are out of equilibrium (see Paper~II). Indeed, this is because the DM density may not have suffered from the Galactic tidal field during less than a single orbit \citep[see their discussion section]{Cardona-Barrero2023}.\\ 

It is beyond the scope of this paper to search for the maximal amount of DM in the Sculptor progenitor to allow a full transformation of its kinematics, because it would require many additional simulations. We noticed that most of the DM content (about 70\%) is not removed from the dwarf after a first passage at pericenter \citep[see their Figure 1]{Mayer2010}. Such a DM removal by Galactic tides is much less efficient than the ram pressure effect on gas that is fully removed. During our simulations of DM-free progenitors, we have noticed that if the initial gas component is more than twice the stellar component within the effective radius, it is sufficient to fully transform the dwarf kinematics after a single pericenter passage. Applying this to the combination of stellar and DM components, it implies an initial DM mass that has to be smaller than the initial gas mass within an effective radius of 279 pc, which is 6.4 $\times 10^5$ M$_{\odot}$  i.e., much smaller than $2.2\times10^7$ M$_{\odot}$ within a 279 pc half-light radius of Sculptor.

 As a consequence, we have followed Paper~II that indicates the difficulty to estimate a dark matter content in dwarf galaxies out of equilibrium. In our models, progenitors of MW dwarfs have been assumed to be made by just a combination of gas and stars, with the goal to verify whether this is sufficient to explain the observations of the three dwarf galaxies considered here. One may also wonder why dIrrs that are known to contain a dominant DM component can be progenitors of dSphs such as Sculptor. This has been investigated by \citet[see their Section 5.2]{Hammer2019} who found that only half of dIrrs selected from {\it Spitzer Photometry and Accurate Rotation Curves} (SPARC, \citealt{Lelli2016}) show evidence for a DM dominant component. 


\section{Simulations of gas-rich dwarfs falling into the Milky Way halo}
\label{sec:simus}
\subsection{Orbits of dwarf progenitors}
\label{sec:orbits}

Besides Sculptor that characterizes most dSphs well, the discovery of Antlia II \citep{Torrealba2019} and Crater II \citep{Torrealba2016} have prompted many questions in the Local Group community. Why do these galaxies have sizes comparable to that of the LMC, while their stellar masses are about 10, 000 times smaller? Why do their DM densities appear to be significantly smaller than that of most dwarfs  \citep{Ji2021,Hayashi2023}? Our simulation resolution is unable to reach very low mass dSphs or ultra-faint dwarfs (see Section~\ref{sec:initial}), but we verify during our preliminary simulations that both Antlia II and Crater II may have similar progenitors as Sculptor, after only changing the gas to stellar mass ratio.

We have generated the orbits for three dwarf galaxies following the {\it Gaia} DR3 proper motions calculated for different MW mass profiles by \citet[see also \citealt{Pace2022}]{Li2021}. From the high mass to the low mass MW model, the pericenters may vary from 54 kpc to 73 kpc (Sculptor), 56 kpc to 102 kpc (Antlia II), and 33 kpc to 58 kpc (Crater II), respectively. Another difficulty in calculating precise orbits is coming from the loss of orbital energy caused by ram pressure exerted on the dwarf before gas exhaustion. \\


Using the above methodology, simulated dwarfs will have positions and velocities that roughly match that from observations. In the following analyses, we have defined coordinate systems centered on each dwarf with orthographic projection in the sky as they are defined from observations \citep{Luri2021}. To compare with observations \citep{Li2021}, we have used the Galactic radial velocity which automatically remove the effect of solar reflex motion. Table~\ref{tab:FitPars} summarizes the comparison with observations, which are described in the three next sections. It includes comparison of orbital quantities and of structural properties. Given the large number of parameters, we cannot expect a perfect match that also change with the choice of the precise epoch. However, it can help us to identify methods for improving the modeling, e.g., by adapting the 3D velocity of the simulated Crater~II (see discussion in Sect.~\ref{sec:Ant-Cra}).\\
Comparison of the 3 simulated dwarf to observations are shown in Table~\ref{tab:FitPars}.

\begin{table*}
\centering
\caption{Comparison between modeling results with observations}
\tabcolsep=3.5pt
    \begin{tabular}{p{3.2cm} p{1.0cm} p{1.7cm} |lll|lll|lll|r}
        \hline\hline
        Parameters         & Symbol & Units & \multicolumn{3}{c}{Sculptor} &
        \multicolumn{3}{c}{Antlia II} & \multicolumn{3}{c}{Crater II}&
        Refs\\[+.1cm]
        \cline{5-6}
        \cline{8-9}
        \cline{11-12}
                           &         &       & &  Obs. & Model  & & Obs. & Model & & Obs. & Model  &\\
        \hline
         RA           & $\alpha$ & deg     && 15.0183    & 14.8570 && 143.8079 & 143.4475 && 177.310 & 178.0151  & (1),(2) \\
         Dec          & $\delta$ & deg     && $-33.7186$   & $-33.8081$ && $-36.6991$ & $-36.4327$ && $-18.413$ & $-18.7613$ &(1),(2) \\
 Galactic radial velocity  & V$_\mathrm{gsr}$&km s$^{-1}$ && $75.9^{+0.7}_{-7.0}$ & 115  && $49.9^{+0.4}_{-0.4}$ & 102  && $-81.4^{+0.3}_{-0.3}$ & $-172$ & (1),(2)   \\
     3D velocity &  $V_{\rm 3D}$& km s$^{-1}$ && $179.6^{+8.3}_{-8.3}$ & 202.2 && $136.9^{+11.7}_{-11.7}$ &  $142.6$ && $124.2^{+12.6}_{-12.1}$ & 241.8   &(1),(2) \\
eccentricity &  $\rm ecc $&  && $0.37-0.52$ & 0.55 && $0.41-0.50$ &  $0.72$  && $0.58-0.60$ & 0.85 &(1),(2)  \\
        \hline
        Helio-distance     & $d_\odot$& kpc  &&  86          & 83    & &    131.8      & 148    & &   117.5      & 100     &  (1),(2)\\
        Effective radius   & $R_{\rm e}$    & kpc   && 0.311        & 0.29   & &     2.54      & 3.98  & &    1.07      & 1.05    &  (3),(2)\\
        Exponential  stellar mass$^a$   & M$_\star$ &10$^6$M$_{\odot}$    &&5.1& 4.27 & &    1.89       &  6.9  & &    0.4       & 0.72    & (4),(2) \\      
Radial velocity gradient & $k_v$ &km s$^{-1}$ kpc$^{-1}$   &&$\sim$ 0.76 & 0.5 & &    2.49       & 1.3 & &    1.07      & 0.7       & (5),(2) \\
  Velocity dispersion$^b$   &$\sigma_\mathrm{los}$& km s$^{-1}$&&8.9&$\sim6.8$ && 5.98 & $\sim 6$   && 2.34  & $\sim 2.5$  & (6),(2)\\
        \hline
     \end{tabular}
    \label{tab:FitPars}
    \parbox{\hsize}{References (last column): 
  (1) \citet{Li2021}; (2) \citet{Ji2021}; (3) \citet{Munoz2018}; (4) \citet{Bettinelli2019}; (5) \citet{Martinez-Garcia2023}; (6) \citet{Iorio2019}. $^a$ ``Exponential'' stellar mass is that derived from the single exponential profile, which would neglect the faint additional component at large radii shown in panels (f) of  Figures~\ref{Scl}, ~\ref{AntII}, and ~\ref{CraII}. $^b$ The velocity dispersion measured for Sculptor analog is $\sigma_\mathrm{los}$, and $\sigma_\mathrm{gsr}$ for Antlia II and Crater II analogs.}
\end{table*}

\subsection{Modeling the Sculptor dSph properties}
\label{sec:Sculptor}
\begin{figure*}
\centering
\includegraphics[width=17cm]{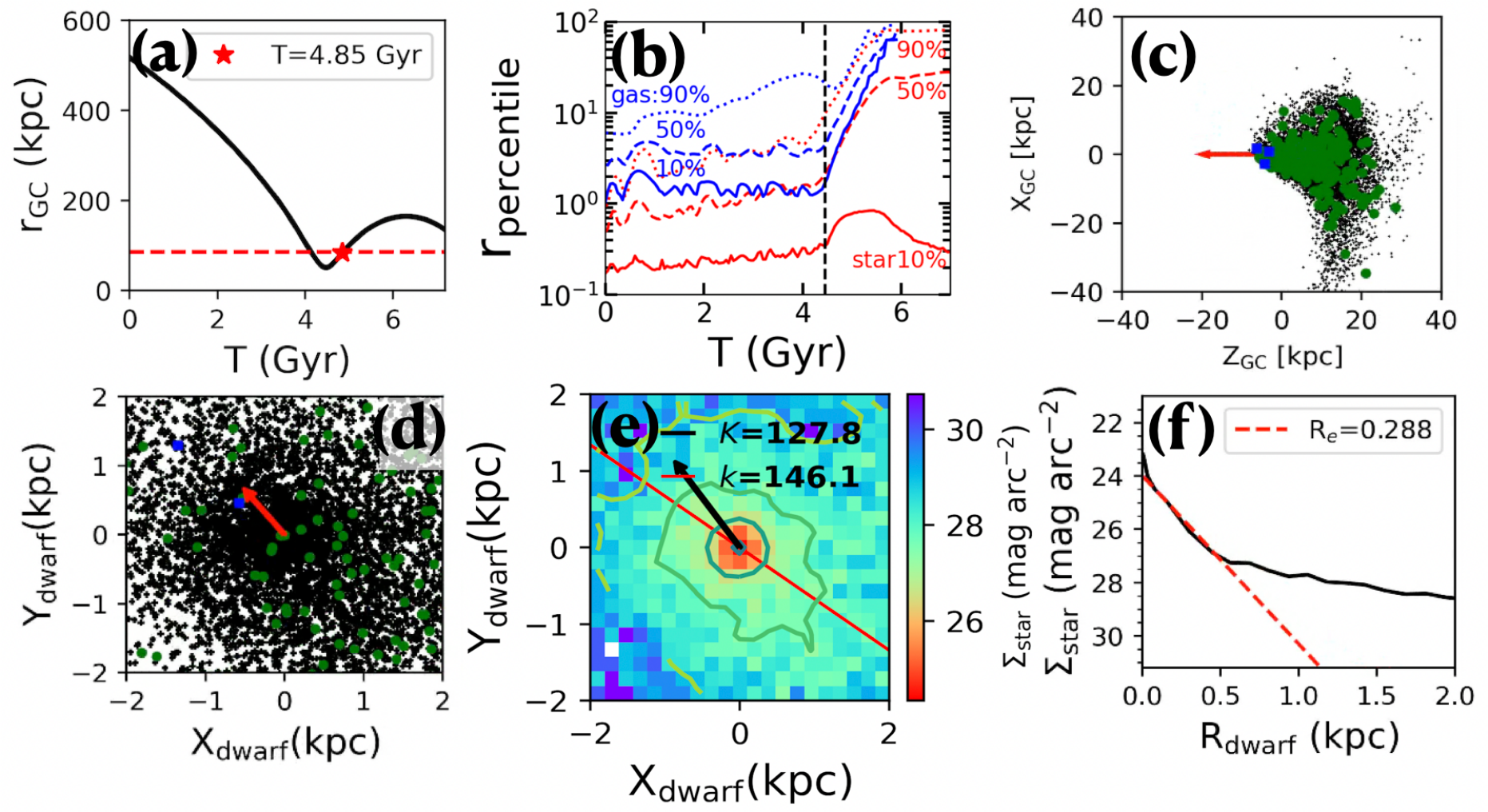}
\caption{Simulation model of a gas-rich dwarf infall into Galactic corona, with the gas being removed close to pericenter. This model reproduces the Sculptor dwarf.
Panel (a): The orbital history of the simulated dwarf (black line), for which the red star indicates the present-day dwarf position. At this time, the gas has been stripped by ram-pressure. 
Panel (b): Evolution of the radii containing 10\% (solid line), 50\% (dashed line) and 90\% (dotted-line) of the total mass, respectively. Red (blue) lines indicate the stellar (gas) component, respectively. The vertical line indicates the time at pericenter passage.
Panel (c): Stellar particle distribution in the orbital plane, for which the Z axis is pointing toward to the Galactic center (see the red arrow). 
Panel (d): Star particle distribution on sky. The red arrow indicates the proper motion direction. Green dots indicate stars that have been formed during the simulation, while the two magenta dots identify very young stars formed 0.5 Gyr ago.
Panel (e): Surface brightness profile overlapping contours showing the stellar shape of the dwarf. The red line shows the major axis. The black arrow indicates the proper motion direction, which is 18.3 degrees offset from the major axis. 
Panel (f): The black line shows the stellar surface brightness profile assuming $M_{\rm stars}/L_{\rm V}=2.5$, while the dashed red line indicates an exponential profile fitting the central region (half-light radius value is indicated on the top-right of the panel). }
\label{Scl}
\end{figure*}

Sculptor is a classical dSph whose properties are reproduced in Figure~\ref{Scl} by our simulations. Panel (a) indicates that the dwarf is just 350 Myr after pericenter, at 83 kpc from the Galactic center, with a position on the sky similar to that observed (see Table~\ref{tab:FitPars}).

Panel (b) of Figure~\ref{Scl} shows how the gas and stellar radii evolve with time. It gives the time evolution of radii including different fractions of gas mass (blue lines) or of stellar mass (red lines), respectively. Dotted, dashed, solid lines correspond to radii including 90\%, 50\%, 10\% of gas/stellar mass, respectively. At pericenter ($T$$\sim 4.5$ Gyr), the gas radius expands rapidly, which indicates that ram-pressure is efficient enough to strip the gas.  The dwarf exhausts its gas 250 Myr later, and then stars are rapidly expanding (see red lines). In the meantime, tidal shocks at pericenter inject energy into the expanding stellar system controlling its 3D motion. Their impact is to accumulate stars along the line of sight (see panel c), which is also the direction of the  MW gravity vector (see also Eq. 11 of Paper II). Both gas loss and tidal shocks boost the velocity dispersion, the former because stars kept their initial velocity dispersion enhanced by the initial gas gravity.  MW tides strongly stretch the object at large scales. Besides this,  stars in the central part of the dwarf move sufficiently fast that they are adiabatically invariant to the action of MW tides, and they present an almost round distribution as indicated by panels (d, e), without evidence for tidal stripping signatures as it is observed within $\Sigma_{\rm stars} <$ 28 mag arcsec$^{-2}$.

Panel (e) indicates a 18.3 degree offset between the proper motion direction and the major axis. The position angle (PA) of Sculptor is 94 degree \citep{McConnachie2020}. The proper motion direction after correcting the solar reflex motion can be derived from {\it Gaia} EDR3 \citep{Li2021}. The observed difference between the PA and proper motion direction is 56.8 degree, which corresponds to a higher misalignment  than in our simulation. 

Panel (f) shows the surface brightness distribution assuming a stellar mass to $V$-luminosity ratio of 2.5. The red-dashed line indicates the exponential function that fits the central region, and this gives a half-light radius of 0.29 kpc, which is consistent with that observed (0.279 kpc, see \citealt{Munoz2018}). There is an additional component at the simulated Sculptor outskirts as it is indicated by the black line, which is consistent with the break observed by \citet[see also \citealt{Yang2022a}]{Westfall2006}. \citet{Sestito2023} found that the surface number density of Sculptor starts to deviate from exponential with an extra component at $R> 0.8$ kpc, which is roughly consistent with our simulation ($R>0.5$ kpc). 

\begin{figure*}
\centering
\includegraphics[width=17cm]{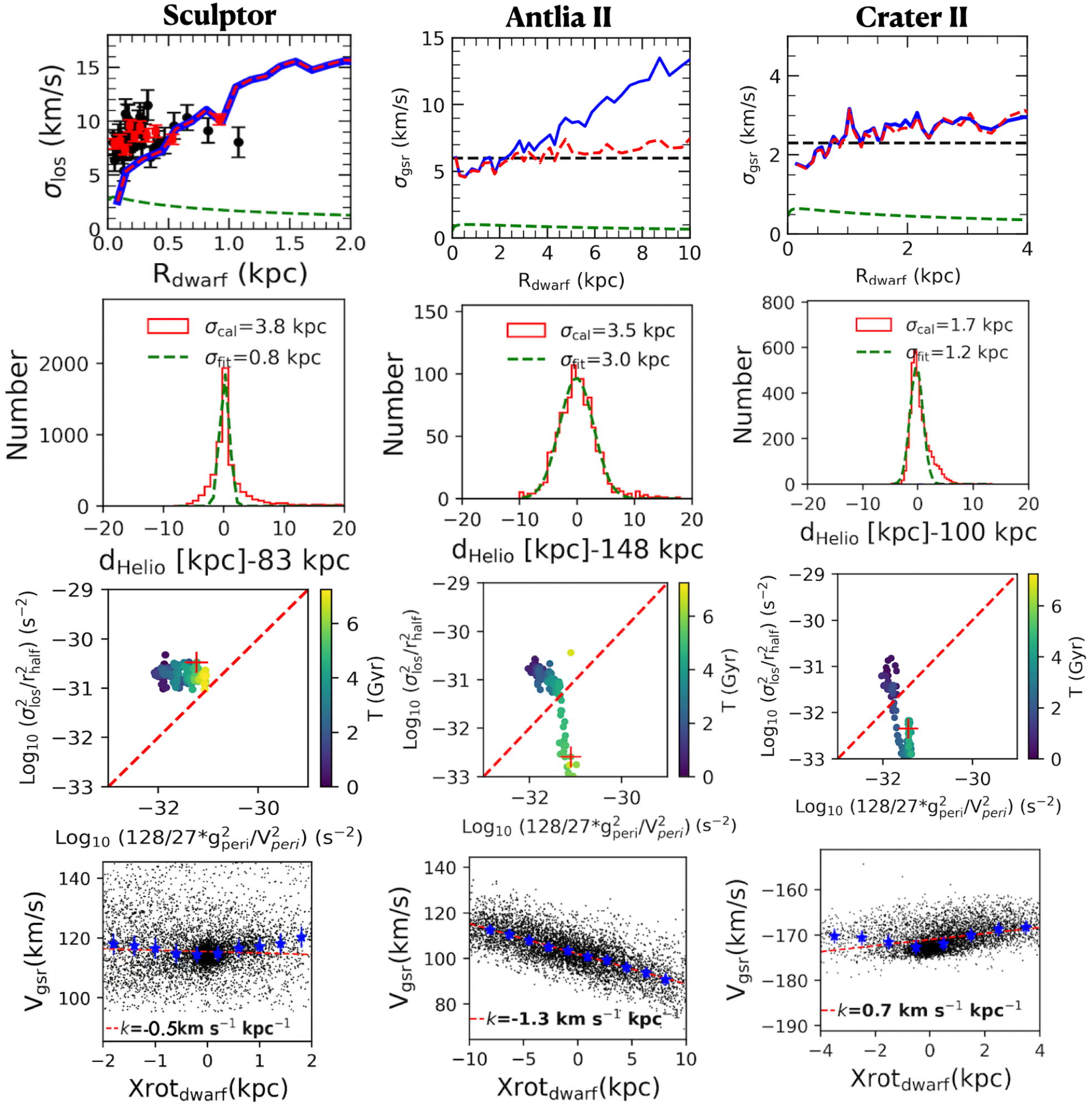}
\caption{Kinematics properties of the 3 dwarf galaxies, Sculptor (first column), Antlia II (second column) and Crater II (third column).
Top panels: Radial velocity dispersion distribution as a function of the radius. The blue line shows the measured values, while the red-dashed line shows results after subtracting the radial velocity gradient as it is done for observed quantities \citep{Walker2009}. The dashed green line shows the contribution to the radial velocity dispersion that is due to the stellar mass (see text). For Sculptor, the observed velocity dispersions at different radii are shown by solid black circles for \citet{Walker2009} and with red dots for \citet{Iorio2019} observations, respectively. 
Second-row panels: Simulated histogram for the distribution of stellar distances (minus the average distance of the dwarf) for stars selected within a 2 kpc radius.
Third-row panels: The relation between $\sigma_{\rm los}^2/r_{\rm half}^2$ and the theoretical value for tidal shocks as it is described in Eq.~\ref{Eq:tidalshocks} (red-dashed line indicate the equality for which tidal shocks provide to the dwarf an energy equals to its binding energy). Red cross represents the present-day position of the dwarf and colored points indicate its evolution during the whole simulation. 
Fourth-row panels: Radial velocity distribution along the major axis. The blue stars and error bars indicate the median values and their uncertainties. The red line shows the gradient value that is given in the bottom left. 
}
\label{kinematics}
\end{figure*}

Fourth row-left panel of Figure~\ref{kinematics} shows a weak velocity gradient in the central region, which could be consistent with the small radial velocity gradient (see Table~\ref{tab:FitPars}) observed within a central area of $\sim 0.73$ kpc, or $\sim$ 0.5 degree \citep[see  their Figure 4]{Martinez-Garcia2023}. However, the observed velocity gradient might be larger at the outskirts ($\sim$ 5 km s$^{-1}$ kpc$^{-1}$, after accounting for the two points at both outskirts of Figure 1 of  \citealt{Battaglia2008}), which are not reproduced by our simulations.

Top-left panel of Figure~\ref{kinematics} shows the simulated velocity dispersion profile compared to that observed, with black and red points representing the observations from \citet{Walker2009} and from
\citet{Iorio2019}, respectively.  The simulated velocity profile matches well the observed value, except in the very inner central part of the dwarf, where the velocity dispersion drops to lower values than that observed. To check how much velocity dispersion is contributed by the stellar mass component, we have built a stellar model with projected surface mass distribution following an exponential function based on panel (f) of Figure~\ref{Scl}. By assuming an isotropic velocity dispersion caused by the stellar mass, we can derive the line of sight velocity dispersion against the radius, which is shown by the green dashed line in the top-left panel of Figure~\ref{kinematics}. The stellar mass contribution to the velocity dispersion is much lower than that observed. Consequently, gas removal and tidal shocks are responsible for the observed large velocity dispersion, and this stands for a duration of about 0.35 Gyr. 

Panel (d) of Figure~\ref{Scl} shows the on-sky stellar particles distribution. Magenta points identify very young stars (age less than 0.5 Gyr), which have been formed just before the gas removal. They are also shown in panel (c) together with green points that represents stars formed during the whole simulation. Strong ram-pressure compress the gas before removing it, which helps to form new stars. 

The third row-left panel of Figure~\ref{kinematics} is based on Eq. 8 and Figure 5 of Paper II, i.e., the relation between $\sigma_{\rm los}^2/r_{\rm half}^2$ and the  strength of the tidal effect at pericenter : 
\begin{equation}
\label{Eq:tidalshocks}
\frac{\sigma^2}{r_{\rm half}^2} = {128\over 27 } \, \left({g_{\rm MW}(R_{\rm peri}) \over V_{\rm peri}}\right)^2  ,   
\end{equation}
where $g_{\rm MW}(R_{\rm peri})$= $G M_{\rm MW}(R_{\rm peri})/R_{\rm peri}^2$ is the gravitational acceleration
exerted by the assumed spherical MW at $R_{\rm peri}$. Objects for which the right term of Eq.~\ref{Eq:tidalshocks} dominates are known to be tidally stripped (e.g., Sgr and Pal 5, see Figure 5 in Paper II). The simulated Sculptor is above the line, which is consistent with its observed value in Paper II (see their Figure 5).

\subsection{Modeling the Antlia II dwarf galaxy properties}

\begin{figure*}
\centering
\includegraphics[width=17cm]{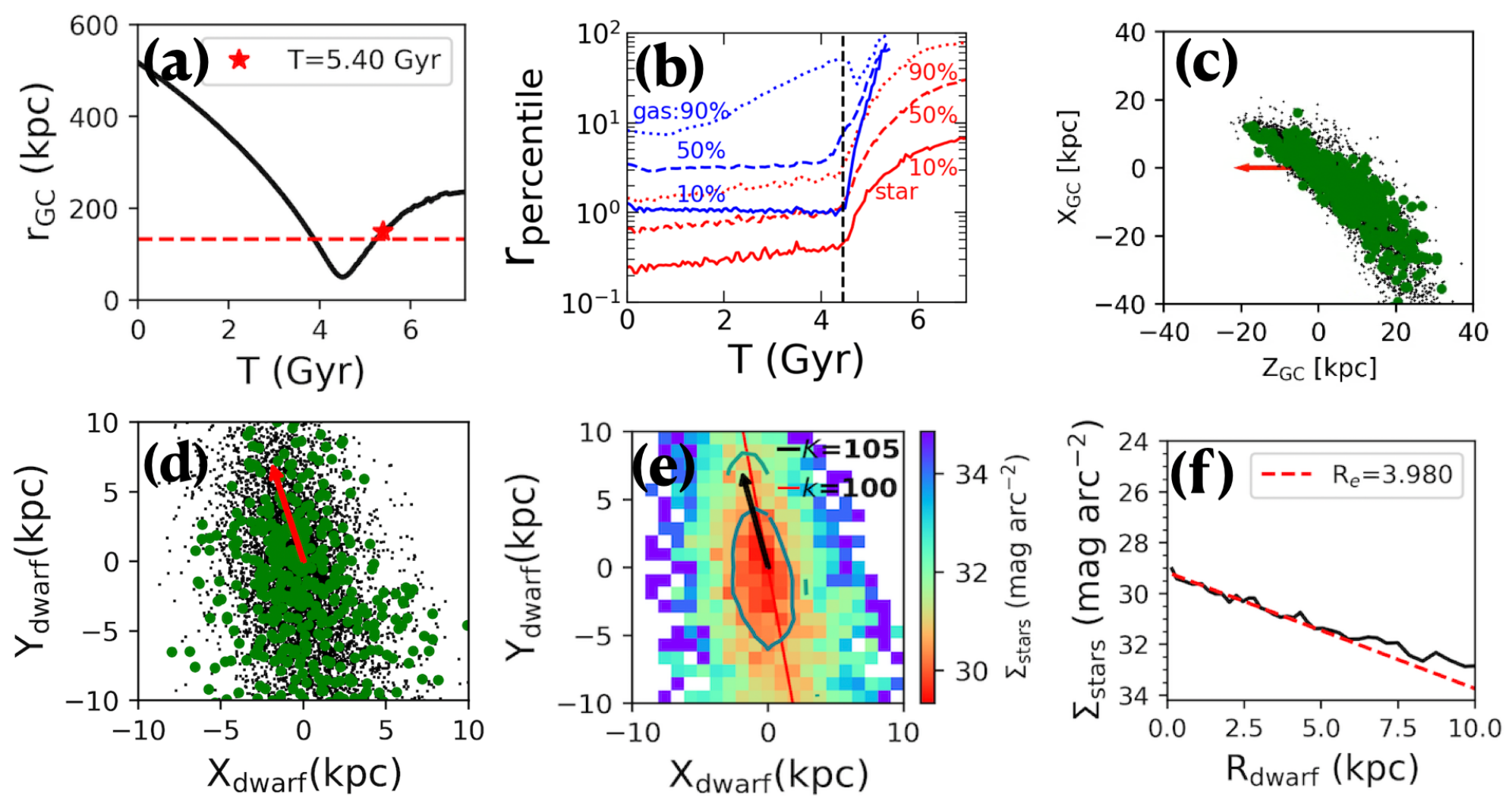}
\caption{Same as Figure~\ref{Scl} but for the Antlia II dwarf.
The description of each panel is similar to that of Figure~\ref{Scl} caption.}
\label{AntII}
\end{figure*}

Antlia II is the only MW dwarf galaxy that has been discovered using {\it Gaia} DR2 proper motions \citep{Torrealba2019}. It possesses a very large radius ($r_{\rm half}= 2.54$ kpc), an extremely low surface brightness even when compared to the most ultra-faint dwarfs, and an ellipticity of 0.6.  Its velocity dispersion is 5.98 km s$^{-1}$, and it shows a large radial velocity gradient (2.49 km s$^{-1}$ kpc$^{-1}$). 

Antlia II has a positive Galactic radial velocity \citep{Li2021,Ji2021} and is leaving its pericenter toward its apocenter.  Figure~\ref{AntII} shows the simulated properties of our modeled Antlia II, which appear quite similar to that of the observed Antlia II. 

Panel (a) of Figure~\ref{AntII} shows the orbital evolution of the dwarf progenitor during its infall into the Galactic halo. Antlia II has lost its gas at pericenter passage ($T$$_\mathrm{peri}=$4.85 Gyr, see Panel b), 0.65 Gyr before its present location (see the red star) at 143 kpc from the Galactic center, which is comparable to the observed value \citep[132 kpc,][]{Ji2021}. 
Panels (c) and (d) shows the stellar particle distribution, for which green dots indicate the stars formed after the beginning of the simulation. Table~\ref{tab:FitPars} indicates that the dwarf is roughly at its current observed position.

Since the simulated Antlia II has left its pericenter where it has suffered strong tidal shocks and gas stripping, the dwarf has been stretched  as shown by the elongated morphology in Panels (c), (d), and (e). In panel (e), the major axis (red line) and the proper motion direction (black arrow, after correcting the solar reflex motion) shares roughly the same direction with a difference of $\sim$ 5 degrees, which is comparable to the observational value ($\sim$ 15 degrees from \citealt{Ji2021}).

Fourth row-middle panel of Figure~\ref{kinematics} shows the line-of-sight velocity distribution that is projected on sky. There is a significant velocity gradient of $\sim 1.3$ km s$^{-1}$ kpc$^{-1}$ along the major axis, which is comparable but a bit smaller than the observed value (2.49 km s$^{-1}$ kpc$^{-1}$, \citealt{Ji2021}). 

Gas loss leads to an expansion of the stellar component, while Galactic tides strongly stretch the simulated dwarf along the orbital direction, as it is illustrated in Panels (c) and (d). Panel (f) shows the surface brightness profile after assuming a stellar mass to light ratio of 2.5. The simulated central surface-brightness is very low ($\Sigma_{\rm stars} \ge $29 mag arcsec$^{-2}$), and consistent with the observed value within the half-light radius (30.7 mag arc$^{-2}$, \citealt{Ji2021}). The half light radius is $\sim 4$ kpc after fitting the profile with an exponential law. It is slightly larger than that observed, for which the major axis radius is around 4 kpc \citep{Ji2021}. Third row-middle panel of Figure~\ref{kinematics} strongly supports the presence of strong tidal shocks and stripping since both the simulated and the observed Antlia II are found with a dominant right-term of Eq.~\ref{Eq:tidalshocks}, i.e., well below the dashed red line (see also Figure 5 of Paper II).

The top-middle panel of Figure~\ref{kinematics} shows the radial profile of the velocity dispersion. The blue line shows the velocity dispersion as a function of radius. The red-dashed line indicates the velocity dispersion after removing the velocity gradient, as it is done in the analysis of the observations \citep{Ji2021}. 
It shows a velocity dispersion between 5 and 7 $\rm km s^{-1}$, 
which is comparable to the observed value for Antlia II \citep[5.98 km s$^{-1}$,][]{Ji2021}.

\subsection{Modeling the Crater II dwarf galaxy properties}

\begin{figure*}
\centering
\includegraphics[width=17cm]{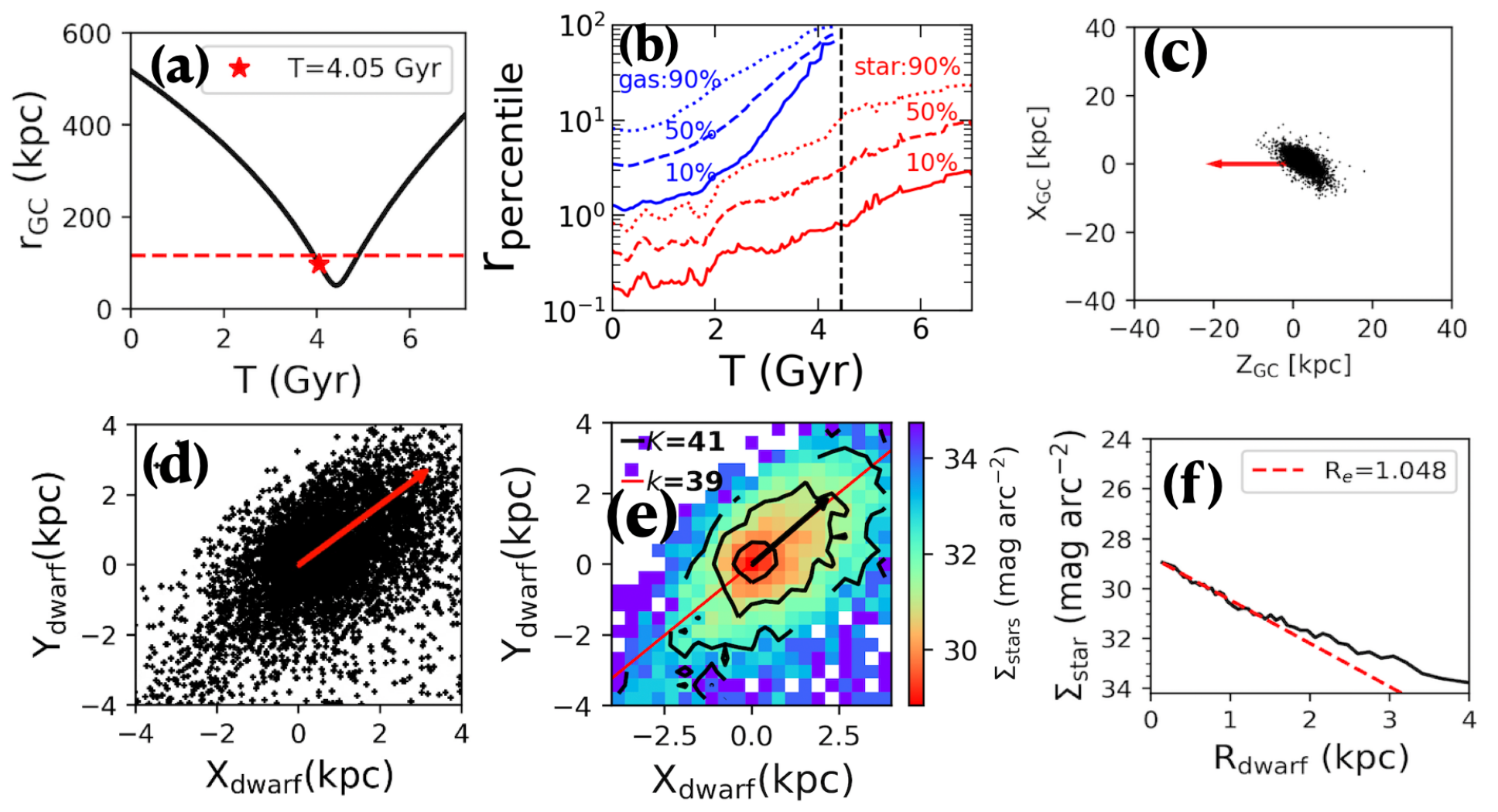}
\caption{Same as Figure~\ref{Scl} but for the Crater II
dwarf 4.05 Gyr after the beginning of the simulation. The description of each panel is similar to that of Figure~\ref{Scl} caption.
}
\label{CraII}
\end{figure*}

Crater II is located at $\sim117$ kpc, and it shows an extended structure with a half-light radius of $\sim$ 1 kpc, and a stellar mass of $\sim 3.5\times 10^{5}$ M$_\odot$ \citep{Ji2021}. Crater II is considered to be difficult to interpret because of its low velocity dispersion ($\sim 2.3$ km s$^{-1}$), especially when it is compared to dwarfs with similar stellar mass. 

In our simulation, the Crater II progenitor has lost its gas content well before pericenter passage. Panel (b) of Figure~\ref{CraII} shows how the gas and stellar radii evolve with time. The sizes of both gas and stellar components rapidly increase during the dwarf infall into the MW halo, due to the gas removal caused by ram-pressure. At $T$ $\sim$ 3.3 Gyr, only 10\% of gas remains within a 10 kpc radius, and none within a 4 kpc radius, which indicates that the gas has been fully removed from the dwarf core. This shows a very efficient gas stripping because of the shallow gravitational potential adopted for the Crater II progenitor.

Gas loss is accompanied by a gravity loss that makes the stellar component expand, as indicated by the red lines of panel (b) of Figure~\ref{CraII}. At $T$=4.05 Gyr, the dwarf arrives at $\sim 100$ kpc from the Sun, with a on-sky position that is roughly consistent with that observed (see Table~\ref{tab:FitPars}).  Due to its expansion, the stellar component becomes very extended with a half-mass radius of $\sim 1$ kpc according to the fit of an exponential profile as shown in the panel (f). In the latter panel the surface brightness distribution has been calculated after assuming a stellar mass to visible light ratio of 2.5. The fitted total mass is $\sim$ 7$\times 10^5$ M$_{\odot}$, which can be compared to the observed value of 3.5 $\times 10^5$ M$_{\odot}$ according to \citet{Ji2021}. Observations may have more difficulty to retrieve the total stellar mass of such a faint object when compared to the idealized case of simulations, and we consider that the simulation result has reached a reasonable agreement with the real Crater II.  

Our modeling predicts large amounts of stars distributed in the far dwarf outskirts due to the inflated stellar component in three dimensions. Panel (c) of Figure~\ref{CraII} shows the stellar particle distribution in Galactocentric coordinate system with Z direction pointing to the Galactic center, confirming the expansion in both Z and X directions. Panel (d) shows the on-sky stellar distribution at large scales, and it confirms the stellar expansion in the X and Y directions.  Panel (f) identifies an extra-component beyond 2 kpc in addition to that of the exponential core. This outer faint component may be extremely difficult to be observed as it becomes very faint ($\Sigma_{\rm stars} \ge$ 32 mag arc$^{-2}$), but perhaps future data from $Gaia$ may help to retrieve it. 

Crater II has not passed its pericenter and given its large distance to the Galactic center, one would not expect strong effects from Galactic tides. This can be seen by comparing panels (c) of Figure~\ref{CraII} to that of Figures~\ref{Scl} and ~\ref{AntII}, which show much larger (tidal) extents of Sculptor and Antlia~II at the faintest levels, in X and Z direction. Panel (e) shows a modest ellipticity of the core contours, which matches pretty well the observations ($e$= 0.12 from Table 1 of \citealt{Ji2021}). It is only at the ultra-faint outskirts that the morphology shows an elliptical shape. Simulations show an angle of $\sim$ 1.7 degrees between the core major axis with the proper motion direction, while \citet{Ji2021} found them almost perpendicular. However, given the modest ellipticity found in both observations and simulations, this difference may be not so dramatic. \\
Both observations and simulations (see fourth row-right panel of Figure~\ref{kinematics}) show a roughly similar velocity gradient along the major axis with values of 1.07 km s$^{-1}$ kpc$^{-1}$ and of 0.7 km s$^{-1}$ kpc$^{-1}$, respectively. Further fine tuning of the progenitor initial orientation may help to improve the simulation of Crater II, in particular its major axis position angle, which is not well reproduced. For example, we may assume a more elliptical distribution of stars in the progenitor, since the dwarf may have kept a memory of its initial morphology,


The top-right panel of Figure~\ref{kinematics} displays the velocity dispersion profile as a function of the projected radius. The blue-solid line gives the measured radial velocity dispersion profile, and the red-dashed line provides the same profile but after removing the velocity gradient. The velocity dispersion (red-dashed line) shows an almost flat dispersion profile with a value of $\sim$ 2.5 km s$^{-1}$. This low velocity dispersion value is comparable to the observed value (2.34 km s$^{-1}$, \citealt{Ji2021}).  The stellar mass component contributes only marginally to the velocity dispersion (see the green-dashed line), conversely to the effect of the gas lost and stellar expansion.

The third row-right panel of Figure~\ref{kinematics} is based on Eq.~\ref{Eq:tidalshocks}, i.e., the relation between $\sigma_{\rm los}^2/r_{\rm half}^2$ and the tidal effect strength at pericenter. Objects below the red-dashed line are know to be tidally stripped (e.g., Sgr and Pal 5), which is also the case for both the simulated and the observed Crater II (see third row-right panel of Figure~\ref{kinematics} and Figure 5 of Paper II, respectively). This means that Crater II is sufficiently close to pericenter to start being tidally shocked.\\
 However, we notice that the properties of the modeled Crater II are discrepant from that observed, especially because panel (a) of Figure~\ref{CraII} shows that the former has a much larger apocenter than the latter ($r_{apo}$$\sim$ 150 kpc from \citealt{Li2021}). Then, conversely to our modeling of Sculptor and Antlia II, we consider that of Crater II to be just an interesting but incomplete attempt.

\subsection{Numerical convergence and gas removal}
\label{sec:convergence}
Appendix~\ref{sec:ICs} describes the stability of dwarf progenitors when left in isolation, using the initial conditions displayed in Table~\ref{tab:IC}. It shows that the dwarf core properties are stable, while the outskirts can be affected by effects such as feedback and star formation.  

The gas removal process depends on several physical parameters, for example, the orbital pericenter distance, the feedback strength, and also the gas distribution in the progenitors. Dwarfs with smaller
pericenter distance would suffer strong ram-pressure stripping and more efficient gas removal, because the Galactic corona density is larger at smaller radii.  Strong feedback also helps to strip the gas (e.g., \citealt{Hammer2015}). Moreover, the initial gas surface density distribution also affects the gas stripping process, i.e., dwarfs with low gas mass surface density would be more easily stripped.

However, we also have to account for an artificial effect that is linked to the mass ratio between particles of the Galactic corona to that of the dwarf gas \citep{Wang2019}, which likely affects the gas stripping efficiency. Appendix~\ref{sec:degenerate} describes how a large particle mass ratio leads to a more efficient gas stripping. It also indicates that a fine tuning of the above parameters (feedback, gas distribution in dwarfs) allows to recover the reproduction of the dwarfs. However, we underline that parameter values shown in Table~\ref{tab:FitPars} may also be slightly dependent on simulation resolution, a problem that hopefully can be solved with much larger resolution in future simulations. For example, Appendix~\ref{sec:degenerate} shows that by improving the resolution by a factor 3, leads us to, e.g., decrease the initial gas mass of the Antlia~II progenitor by a factor 1.5, which renders it even more realistic its position in Figure~\ref{HImass}.

\section{DISCUSSION}
\label{sec:discussion}

We propose here that most MW dwarf galaxies have been recently fully transformed during their recent infall into the MW halo (see Papers I and II). During the infall of their gas-rich progenitors, they are affected by physical processes including gas removal from ram pressure of the MW halo gas, a subsequent expansion of stars due to the lack of gravity, and MW tidal shocks occurring near pericenter passage. The scenario is consistent with observations, which is illustrated in Paper~I (recent infall from {\it Gaia} measurements) and Paper~II (dependency of their velocity dispersion to orbital parameters). Simulations presented in this paper show that the above processes allow to reproduce many properties (see Table~\ref{tab:FitPars}) including the excess of velocity dispersion observed in dwarf galaxies, because they are out of equilibrium objects. They also permit to understand why very different kinds of dwarfs can be reproduced by variations within the same set of physical processes. Moreover, simulations allow to characterize the star formation induced by ram pressure, and to predict the fraction of young stars that can be detected, in order to compare them with investigations made in all classical dwarfs (see Yang et al. 2023, submitted, hereafter Paper~IV).

\subsection{Gas turbulence during its removal and tidal shocks increase velocity dispersion}

Our dwarf progenitors are fully dominated by the gas component, which imposes very large initial velocity dispersions compared to what would be the contribution of the stellar mass only. It implies that the gas removal leads to a full transformation of the dwarf and of its kinematics. Physical effects are numerous, and they all lead to large values of the line-of-sight velocity dispersion. Immediately after the gas removal, stars expand while keeping their initial (large) velocity dispersion, while the mutual interaction between the gas in the dwarf and in the Galactic corona has created various hydrodynamic phenomena, e.g., shock, compression, or in other words, turbulence.  In the meantime, feedback from supernova explosions of the newly formed stars also adds turbulence to the gas, which affects the stellar component until the gas is removed. Gas turbulence may shake the stellar component, providing an additional source for increasing kinetic energy of dwarf stars. 
Stellar feedback can drive significant kinematic fluctuations of both gas and stars, and this has been extensively studied by other hydrodynamical simulations of dwarf galaxies \citep[e.g.,][]{ElBadry2016,ElBadry2017}.

This has to be added to the impact of tidal shocks, which is especially efficient when the dwarf is found near its pericenter (see paper II). Panels (c) of Figures~\ref{Scl}, ~\ref{AntII}, and ~\ref{CraII} show the projection of simulated stars along the Z axis, towards the MW. It confirms that star expansion and tidal shocks help to accumulate stars along the Z axis towards the MW, which coincides with that of the MW gravity at pericenter ($g_{\rm MW}$).  According to Paper II (see their Eq. 12) it leads to an increase of the square of the velocity dispersion by $\Delta \sigma^2$:
 \begin{equation}
\Delta \sigma^2 = 2\sqrt{2} \, g_{\rm MW} \, r_{\rm half}  \, f_{\rm ff},
\label{Eq:ffs}
 \end{equation}
where $f_{\rm ff}$ (ff standing for free-fall) represents the fraction of stars that have expanded so much that they are no more affected by the dwarf self-gravity, but by the MW gravity, and it depends also on the structural properties of the dwarf (see details in Paper II). According to Paper II, the relation shown in Eq.~\ref{Eq:ffs} between intrinsic parameters and the MW gravity is caused by tidal shocks and gas removal and it leads to a significant correlation shown in their figure 9. \\
One may also wonder whether or not observers can distinguish between stars at different distances from us. For example, in the simulated Sculptor dwarf more than 90\% of stars are within $\pm$ 5 kpc from the core at 83 kpc, according to the second row-left panel of Figure~\ref{kinematics} . Given the expected accuracy of individual star measurements ($\sim$ 20\%), it appears likely that during observational campaigns \citep{Walker2007,Walker2009}, all stars used to probe kinematics have been selected within that distance range.

\subsection{Simulations predict outer stellar halos of dwarf galaxies}

Recent studies have shown that many dwarfs possess a  very extended stellar halo. This includes Tucana II \citep{Chiti2023}, Ursa Minor \citep{Sestito2023}, Bootes I \citep{Longeard2022}, Antlia II \citep{Ji2021},
Fornax \citep{Yang2022a}, Coma Berenices, and Ursa Major \citep{Waller2023}. Although Antlia II shows apparent tidal features, in most other cases, studies have found outskirt stars in all sky directions. \citet{Waller2023} proposed that such stars may have been ejected due to very early and strong feedback. However, \citet{Chiti2023} mentioned that there is no evidence for such an energetic supernova event from their detailed chemical analysis of Tucana II. They suggested a possible early merger, as also proposed by \citet[see also \citealt{Goater2023}]{Sestito2023} for explaining stars found up to 12 half-light radii in the Ursa Minor envelope. 

In our scenario, there are two different mechanisms that lead to expand the stellar component. The main channel to explain stars in the outer envelopes of most dwarfs is the (spherical) expansion of the stellar component after gas loss, i.e., due to the subsequent lack of gravity. This mechanism reproduces well the fact that the observed outskirt stars are found in many directions from the dwarf cores \citep{Chiti2023,Sestito2023,Longeard2022,Yang2022a,Waller2023}. It predicts that stars are expanding in an almost spherical geometry, and without a velocity gradient.
Another mechanism is the tidal stretching, which applies well to Antlia II and perhaps also to Crater II \citep{Ji2021}. However, these two dwarfs are very exceptional by their large size associated to their extremely low surface brightness. Together with Sgr, Antlia II and Crater II are the only dwarfs that are clearly dominated by tidal stripping according to their locations in Figure 5 of Paper II. 
\subsection{Star formation history}

Gas-rich progenitors of MW dwarf galaxies have their gas content pressurized by the MW halo gas, which predicts some star formation during their infall in the MW halo. Consequently there should be some young stars in MW dwarf galaxies, and especially in the most massive ones. In our simulations, this applies to Sculptor and Antlia II, but not to Crater II because of its very low initial gas surface density (compare the green-dots between panels (d) of Figures~\ref{Scl}, ~\ref{AntII}, and ~\ref{CraII}). However, the newly formed stars are affected by gas turbulence, especially at the most recent epochs when only small fractions of gas are still attached to the dwarf core. \\

Turbulent motions lead to quite chaotic motions of recently formed stars, and indeed we find only two stellar particles with young ages ($\le$ 0.5 Gyr) in the core ($\le$ 2 kpc) of the simulated Sculptor dwarf. This corresponds to only 0.2\% of the Sculptor core stellar mass, which could be compared to observations (see Paper~IV). 


\subsection{The difference between Antlia II and Crater II: two extended dwarfs with 
different velocity dispersions}
\label{sec:Ant-Cra}
\begin{figure*}
\centering
\includegraphics[width=17cm]{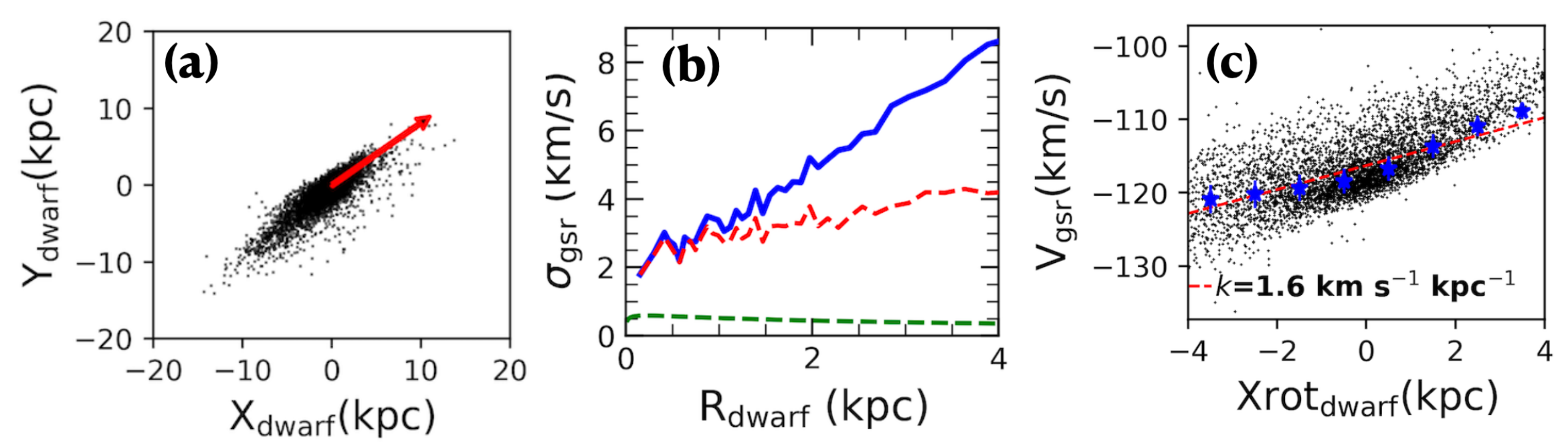}
\caption{The simulated dwarf galaxy Crater II when it will be at pericenter. Panel (a) shows the stellar particle distribution on the sky. Panel (b) shows
the velocity dispersion profile in function of the radius. Panel (c) denotes the
radial velocity distribution along the major axis.}
\label{CraII_later}
\end{figure*}

The major difference between Antlia II and Crater II is related to their different orbital and gas loss histories. Our Crater II progenitor possesses a very low-density gas halo that is removed before the pericenter passage, while the Antlia II gas is removed at pericenter passage. It implies that the kinematics of Antlia II are affected by both gas removal and tidal shocks, as it is observed after pericenter, while those of Crater II are almost exclusively affected by gas removal. This leads to a larger velocity dispersion for Antlia II compared to Crater II, which is also supported by the fact that the former is closer to its pericenter than the latter.  Figure~\ref{CraII_later} shows how Crater II would evolve in the near future, i.e., when arriving at a position close to the pericenter. Panel (a) of Figure~\ref{CraII_later} shows a much more elongated galaxy than that of panel (d) of Figure~\ref{CraII}, i.e., Crater II is likely at an early stage of the tidal stripping (or stretching) process, especially when compared with Antlia II. Panel (b) of Figure~\ref{CraII_later} also indicates a larger velocity dispersion when compared to the top-right panel of Figure~\ref{kinematics}.

The above illustrates the roles of the gas removal and tidal shocks at pericenter, which, when combined, are very efficient in increasing the velocity dispersion. However, our simulations show also some limitations especially for reproducing the Crater II properties. While panels (a) of Figures~\ref{Scl}, and ~\ref{AntII} indicate that both Sculptor and Antlia II become bound to the MW potential, Crater II seems to remain unbound, probably because the gas-removal has been too fast to efficiently slowdown its 3D velocity. It results that while 3D velocities and apocenters of the two first galaxies are well reproduced, the predicted values for Crater II (see Table~\ref{tab:FitPars}) would require further fine-tuning.

Another interest in comparing Antlia II and Crater II is coming from a recent study by \citet{Taibi2023} showing that most dwarfs belonging to the Vast Polar Structure (VPOS, see \citealt{Pawlowski2012}) surrounding the MW are found near but before their pericenters. Since Crater II belongs to the VPOS conversely to Antlia II, one may wonder the origin of such a property for most VPOS dwarfs. Comparing panels (b) of Figures~\ref{AntII} and~\ref{CraII} indicates that after the present time, stars of the former are going to expand very rapidly, which might correspond to a full dissolution of the dwarf in the MW halo. One may wonder whether some of the VPOS dwarfs that have passed their pericenter might share a similar fate to that of the Antlia II.

\subsection{The difference between Sculptor and Antlia~II:  dwarfs with different sizes and with high velocity dispersions }

Both Sculptor and Antlia II have lost their gas near pericenter, a position they have both passed recently. Antlia II's morphology is fully tidally stretched while the Sculptor core does not show significant tidal effects. This is because the core of Sculptor is about 100 times denser than that of Antlia II, as also indicated from the 5 magnitude difference in central surface-brightness. It implies that the former is dense enough to resist the Galactic tides, conversely to the later. This is the reason why Sculptor shows no signature of tides in the central region ($\sim 2$ kpc). However, at larger scales ($r\sim 10$ kpc), the model does predict tidal features. Confirming these would require very deep observations.


The above also indicates another possible limitation of our modeling. For example, the top-left panel of Figure~\ref{kinematics}  shows that in the very central region of Sculptor (within 0.1 kpc) there is a drop
of the modeled velocity dispersion, which appears in contradiction with the observations (see the data points). First, this decreasing velocity dispersion in the center may be caused by the artificial softening effect in the numerical simulation. The softening radius for stellar particles is 0.02 kpc, meaning that the central 0.06 kpc is dominated by artificial softening effects. It lets however unexplained the region between 0.06 to 0.1 kpc\footnote{One may also suspect that because of the expansion of the stellar particles, those seen at 0.1 kpc after the simulation were initially lying in a smaller region and could be also affected by softening effects.}. Further fine-tuning would be necessary to solve this issue, perhaps by considering a slightly less dense core for Sculptor. A possible alternative might come from observations, since the  center of Sculptor may slightly differ from spectroscopic to photometric observations.


\section{CONCLUSIONS}
\label{sec:conclusion}
Analyses of {\it Gaia} DR3 proper motions have shown that most dwarf galaxies are less bound to the MW than are stars or globular clusters associated to the Gaia-Sausage-Enceladus event and to the infall of the Sgr dwarf system, which have occurred 9$\pm$1 and 5$\pm$1 Gyr ago, respectively. Paper I found a linear relation between the infall time and the logarithm of the binding energy, which extrapolated to most MW dwarf galaxies, fell in more recently than 3 Gyr ago. During this elapsed time, most dwarf galaxies would not have time to experience more than one orbit, which suggests that they can be at first infall, as has been proposed by  \citet{Kallivayalil2013} for the Magellanic Clouds. As a consequence, most dwarf progenitors should have been gas-rich at relatively recent lookback times, because most dwarfs at large distances from a giant spiral in the Local Group are gas rich. It opens the possibility that dwarf progenitors have lost their gas recently near their first passage to pericenter, a location where many dwarfs are currently found \citep{Fritz2019,Li2021}. \\

 In Paper~II, we have shown that pericenter passages are also accompanied by tidal shocks, which lead to correlations between orbital and structural properties of dwarfs. The goal of the present Paper~III is to test the above scenario and to verify whether it is consistent with the observed properties of dwarf galaxies. Here, we have considered the complex transformation of gas-rich dwarfs interacting with the MW hot corona (ram pressure) and gravitational field (tides). Simulations have been performed using the GIZMO hydrodynamical solver \citep{Hopkins2015}. There are some limitations for these simulations, which are linked to the different gas particles mass between the MW hot corona and the dwarf galaxy progenitors. This has led us to focus on the specific case of Sculptor, which is massive enough to avoid the above limitations, while it is well representative of classical dwarf spheroidal galaxies. \\
 
 Hydrodynamical simulations of a gas-rich dwarf galaxy falling into the MW halo have been able to reproduce almost all the observed properties of Sculptor, including its large velocity dispersion when compared to expectations from the gravity due to its sole stellar content. The simulations also succeed to reproduce its core morphology, luminosity profile, velocity gradient, and location in the plane relating structural properties and orbital properties linked to the MW gravitational field. In this scenario, Sculptor's stellar core can survive the pericenter passage and gas loss, thanks to its sufficiently large density that allows an adiabatic contraction instead of an expansion that affects many stars in the outskirts. This scenario predicts that most MW dwarf galaxies should have lost stars in their outskirts, which are mostly expanding following a spherical symmetry, reproducing observations of stellar halos for many of them \citep{Sestito2023,Waller2023,Chiti2023,Cantu2021,Yang2022b,Roederer2023}. However, it also predicts that there should be a tiny fraction of very young stars in Sculptor, for which Paper~IV claims their detection.\\
 
 Using a similar orbital path and more gas-rich progenitors, our simulations have also been able to reproduce the properties of two unusual dwarf galaxies, Antlia II and Crater II. Their progenitors are rarer than that of Sculptor according to statistical analyses of local field galaxies, which is consistent with the fact that they are both unusual inhabitants of the Galactic halo. Our simulations explain why Antlia II has a much larger velocity dispersion than Crater II, because the former has passed its pericenter and is strongly affected by MW tidal shocks, while the latter is only affected by stellar expansion caused by the lack of gravity after the gas removal. \\
 
Here, the proposed scenario of ram-pressure stripping and Galactic tidal shocks has begun to be validated by hydrodynamical simulations, providing considerable changes in our understanding of MW dwarf galaxies. Most of the dwarfs are likely out-of-equilibrium, which prevents any measurements of their mass from assuming self-equilibrium conditions. It also provides an elegant and simple solution for understanding the exceptional behavior of the two extremely low surface brightness dwarfs, Antlia II and Crater II. Finally, further simulations could be useful to verify which fractions the DM could reach while remaining consistent with the transformation in a single pericenter passage of a gas-rich, rotation-dominated dwarfs into a gas-free, dispersion-dominated dwarf as the MW dwarfs are observed to be.

\section*{Acknowledgments}

We are very grateful to Piercarlo Bonifacio, Elisabetta Caffau, Yongjun Jiao, and Hefan Li for their participations to the numerous and lively meetings during which this paper has been discussed. We warmly thank the referee for the very useful remarks, which have considerably help the writing of the manuscript.
J.-L.W. acknowledges financial support from the China Scholarship Council (CSC) No.202210740004. Marcel S. Pawlowski acknowledges funding of a Leibniz-Junior Research Group
 (project number J94/2020). We are grateful for the support
of the International Research Program Tianguan, which is an agreement between
the CNRS in France, NAOC, IHEP, and the Yunnan Univ. in China.

\section*{DATA AVAILABILITY}

The data underlying this article will be shared on request
to the corresponding authors.

\bibliographystyle{mnras}
\bibliography{reference.bib}

\providecommand{\noopsort}[1]{}\providecommand{\singleletter}[1]{#1}%
\begin{thebibliography}{}
\makeatletter
\relax
\def\mn@urlcharsother{\let\do\@makeother \do\$\do\&\do\#\do\^\do\_\do\%\do\~}
\def\mn@doi{\begingroup\mn@urlcharsother \@ifnextchar [ {\mn@doi@}
  {\mn@doi@[]}}
\def\mn@doi@[#1]#2{\def\@tempa{#1}\ifx\@tempa\@empty \href
  {http://dx.doi.org/#2} {doi:#2}\else \href {http://dx.doi.org/#2} {#1}\fi
  \endgroup}
\def\mn@eprint#1#2{\mn@eprint@#1:#2::\@nil}
\def\mn@eprint@arXiv#1{\href {http://arxiv.org/abs/#1} {{\tt arXiv:#1}}}
\def\mn@eprint@dblp#1{\href {http://dblp.uni-trier.de/rec/bibtex/#1.xml}
  {dblp:#1}}
\def\mn@eprint@#1:#2:#3:#4\@nil{\def\@tempa {#1}\def\@tempb {#2}\def\@tempc
  {#3}\ifx \@tempc \@empty \let \@tempc \@tempb \let \@tempb \@tempa \fi \ifx
  \@tempb \@empty \def\@tempb {arXiv}\fi \@ifundefined
  {mn@eprint@\@tempb}{\@tempb:\@tempc}{\expandafter \expandafter \csname
  mn@eprint@\@tempb\endcsname \expandafter{\@tempc}}}

\bibitem[\protect\citeauthoryear{{Battaglia}, {Helmi}, {Tolstoy}, {Irwin},
  {Hill}  \& {Jablonka}}{{Battaglia} et~al.}{2008}]{Battaglia2008}
{Battaglia} G.,  {Helmi} A.,  {Tolstoy} E.,  {Irwin} M.,  {Hill} V.,
  {Jablonka} P.,  2008, \mn@doi [ApJL] {10.1086/590179}, \href
  {https://ui.adsabs.harvard.edu/abs/2008ApJ...681L..13B} {681, L13}

\bibitem[\protect\citeauthoryear{{Bettinelli}, {Hidalgo}, {Cassisi},
  {Aparicio}, {Piotto}, {Valdes}  \& {Walker}}{{Bettinelli}
  et~al.}{2019}]{Bettinelli2019}
{Bettinelli} M.,  {Hidalgo} S.~L.,  {Cassisi} S.,  {Aparicio} A.,  {Piotto} G.,
   {Valdes} F.,   {Walker} A.~R.,  2019, \mn@doi [MNRAS]
  {10.1093/mnras/stz1679}, \href
  {https://ui.adsabs.harvard.edu/abs/2019MNRAS.487.5862B} {487, 5862}

\bibitem[\protect\citeauthoryear{{Cantu} et~al.,}{{Cantu}
  et~al.}{2021}]{Cantu2021}
{Cantu} S.~A.,  et~al., 2021, \mn@doi [\apj] {10.3847/1538-4357/ac0443}, \href
  {https://ui.adsabs.harvard.edu/abs/2021ApJ...916...81C} {916, 81}

\bibitem[\protect\citeauthoryear{{Cardona-Barrero}, {Battaglia}, {Nipoti}  \&
  {Di Cintio}}{{Cardona-Barrero} et~al.}{2023}]{Cardona-Barrero2023}
{Cardona-Barrero} S.,  {Battaglia} G.,  {Nipoti} C.,   {Di Cintio} A.,  2023,
  \mn@doi [\mnras] {10.1093/mnras/stad1138}, \href
  {https://ui.adsabs.harvard.edu/abs/2023MNRAS.522.3058C} {522, 3058}

\bibitem[\protect\citeauthoryear{{Catinella} et~al.,}{{Catinella}
  et~al.}{2010}]{Catinella2010}
{Catinella} B.,  et~al., 2010, \mn@doi [MNRAS]
  {10.1111/j.1365-2966.2009.16180.x}, \href
  {https://ui.adsabs.harvard.edu/abs/2010MNRAS.403..683C} {403, 683}

\bibitem[\protect\citeauthoryear{{Chiti} et~al.,}{{Chiti}
  et~al.}{2023}]{Chiti2023}
{Chiti} A.,  et~al., 2023, \mn@doi [The Astronomical Journal]
  {10.3847/1538-3881/aca416}, \href
  {https://ui.adsabs.harvard.edu/abs/2023AJ....165...55C} {165, 55}

\bibitem[\protect\citeauthoryear{{El-Badry}, {Wetzel}, {Geha}, {Hopkins},
  {Kere{\v{s}}}, {Chan}  \& {Faucher-Gigu{\`e}re}}{{El-Badry}
  et~al.}{2016}]{ElBadry2016}
{El-Badry} K.,  {Wetzel} A.,  {Geha} M.,  {Hopkins} P.~F.,  {Kere{\v{s}}} D.,
  {Chan} T.~K.,   {Faucher-Gigu{\`e}re} C.-A.,  2016, \mn@doi [ApJ]
  {10.3847/0004-637X/820/2/131}, \href
  {https://ui.adsabs.harvard.edu/abs/2016ApJ...820..131E} {820, 131}

\bibitem[\protect\citeauthoryear{{El-Badry}, {Wetzel}, {Geha}, {Quataert},
  {Hopkins}, {Kere{\v{s}}}, {Chan}  \& {Faucher-Gigu{\`e}re}}{{El-Badry}
  et~al.}{2017}]{ElBadry2017}
{El-Badry} K.,  {Wetzel} A.~R.,  {Geha} M.,  {Quataert} E.,  {Hopkins} P.~F.,
  {Kere{\v{s}}} D.,  {Chan} T.~K.,   {Faucher-Gigu{\`e}re} C.-A.,  2017,
  \mn@doi [ApJ] {10.3847/1538-4357/835/2/193}, \href
  {https://ui.adsabs.harvard.edu/abs/2017ApJ...835..193E} {835, 193}

\bibitem[\protect\citeauthoryear{{Fritz}, {Battaglia}, {Pawlowski},
  {Kallivayalil}, {van der Marel}, {Sohn}, {Brook}  \& {Besla}}{{Fritz}
  et~al.}{2018}]{Fritz2018}
{Fritz} T.~K.,  {Battaglia} G.,  {Pawlowski} M.~S.,  {Kallivayalil} N.,  {van
  der Marel} R.,  {Sohn} S.~T.,  {Brook} C.,   {Besla} G.,  2018, \mn@doi
  [Astronomy and Astrophysics] {10.1051/0004-6361/201833343}, \href
  {https://ui.adsabs.harvard.edu/abs/2018A&A...619A.103F} {619, A103}

\bibitem[\protect\citeauthoryear{{Fritz}, {Carrera}, {Battaglia}  \&
  {Taibi}}{{Fritz} et~al.}{2019}]{Fritz2019}
{Fritz} T.~K.,  {Carrera} R.,  {Battaglia} G.,   {Taibi} S.,  2019, \mn@doi
  [\aap] {10.1051/0004-6361/201833458}, \href
  {https://ui.adsabs.harvard.edu/abs/2019A&A...623A.129F} {623, A129}

\bibitem[\protect\citeauthoryear{{Gaia Collaboration} et~al.,}{{Gaia
  Collaboration} et~al.}{2021}]{Luri2021}
{Gaia Collaboration} et~al., 2021, \mn@doi [A\&A]
  {10.1051/0004-6361/202039588}, \href
  {https://ui.adsabs.harvard.edu/abs/2021A&A...649A...7G} {649, A7}

\bibitem[\protect\citeauthoryear{{Goater} et~al.,}{{Goater}
  et~al.}{2023}]{Goater2023}
{Goater} A.,  et~al., 2023, \mn@doi [arXiv e-prints]
  {10.48550/arXiv.2307.05130}, \href
  {https://ui.adsabs.harvard.edu/abs/2023arXiv230705130G} {p. arXiv:2307.05130}

\bibitem[\protect\citeauthoryear{{Grcevich} \& {Putman}}{{Grcevich} \&
  {Putman}}{2009}]{Grcevich2009}
{Grcevich} J.,  {Putman} M.~E.,  2009, \mn@doi [ApJ]
  {10.1088/0004-637X/696/1/385}, \href
  {http://adsabs.harvard.edu/abs/2009ApJ...696..385G} {696, 385}

\bibitem[\protect\citeauthoryear{{Hammer}, {Yang}, {Flores}, {Puech}  \&
  {Fouquet}}{{Hammer} et~al.}{2015}]{Hammer2015}
{Hammer} F.,  {Yang} Y.~B.,  {Flores} H.,  {Puech} M.,   {Fouquet} S.,  2015,
  \mn@doi [ApJ] {10.1088/0004-637X/813/2/110}, \href
  {http://cdsads.u-strasbg.fr/abs/2015ApJ...813..110H} {813, 110}

\bibitem[\protect\citeauthoryear{{Hammer}, {Yang}, {Wang}, {Arenou}, {Puech},
  {Flores}  \& {Babusiaux}}{{Hammer} et~al.}{2019}]{Hammer2019}
{Hammer} F.,  {Yang} Y.,  {Wang} J.,  {Arenou} F.,  {Puech} M.,  {Flores} H.,
  {Babusiaux} C.,  2019, \mn@doi [\apj] {10.3847/1538-4357/ab36b6}, \href
  {https://ui.adsabs.harvard.edu/abs/2019ApJ...883..171H} {883, 171}

\bibitem[\protect\citeauthoryear{{Hammer} et~al.,}{{Hammer}
  et~al.}{2023}]{Hammer2023}
{Hammer} F.,  et~al., 2023, \mn@doi [MNRAS] {10.1093/mnras/stac3758}, \href
  {https://ui.adsabs.harvard.edu/abs/2023MNRAS.519.5059H} {519, 5059}

\bibitem[\protect\citeauthoryear{Hammer et~al.,}{Hammer
  et~al.}{2024}]{Hammer2024}
Hammer F.,  et~al., 2024, \mn@doi [Monthly Notices of the Royal Astronomical
  Society] {10.1093/mnras/stad2922}, 527, 2718

\bibitem[\protect\citeauthoryear{{Hayashi}, {Hirai}, {Chiba}  \&
  {Ishiyama}}{{Hayashi} et~al.}{2023}]{Hayashi2023}
{Hayashi} K.,  {Hirai} Y.,  {Chiba} M.,   {Ishiyama} T.,  2023, \mn@doi [\apj]
  {10.3847/1538-4357/ace33e}, \href
  {https://ui.adsabs.harvard.edu/abs/2023ApJ...953..185H} {953, 185}

\bibitem[\protect\citeauthoryear{{Higgs}, {McConnachie}, {Annau}, {Irwin},
  {Battaglia}, {C{\^o}t{\'e}}, {Lewis}  \& {Venn}}{{Higgs}
  et~al.}{2021}]{Higgs2021}
{Higgs} C.~R.,  {McConnachie} A.~W.,  {Annau} N.,  {Irwin} M.,  {Battaglia} G.,
   {C{\^o}t{\'e}} P.,  {Lewis} G.~F.,   {Venn} K.,  2021, \mn@doi [MNRAS]
  {10.1093/mnras/stab002}, \href
  {https://ui.adsabs.harvard.edu/abs/2021MNRAS.503..176H} {503, 176}

\bibitem[\protect\citeauthoryear{{Hopkins}}{{Hopkins}}{2015}]{Hopkins2015}
{Hopkins} P.~F.,  2015, \mn@doi [MNRAS] {10.1093/mnras/stv195}, \href
  {http://cdsads.u-strasbg.fr/abs/2015MNRAS.450...53H} {450, 53}

\bibitem[\protect\citeauthoryear{{Huang}, {Haynes}, {Giovanelli}  \&
  {Brinchmann}}{{Huang} et~al.}{2012}]{Huang2012}
{Huang} S.,  {Haynes} M.~P.,  {Giovanelli} R.,   {Brinchmann} J.,  2012,
  \mn@doi [ApJ] {10.1088/0004-637X/756/2/113}, \href
  {https://ui.adsabs.harvard.edu/abs/2012ApJ...756..113H} {756, 113}

\bibitem[\protect\citeauthoryear{{Iorio}, {Nipoti}, {Battaglia}  \&
  {Sollima}}{{Iorio} et~al.}{2019}]{Iorio2019}
{Iorio} G.,  {Nipoti} C.,  {Battaglia} G.,   {Sollima} A.,  2019, \mn@doi
  [MNRAS] {10.1093/mnras/stz1342}, \href
  {https://ui.adsabs.harvard.edu/abs/2019MNRAS.487.5692I} {487, 5692}

\bibitem[\protect\citeauthoryear{{Ji} et~al.,}{{Ji} et~al.}{2021}]{Ji2021}
{Ji} A.~P.,  et~al., 2021, \mn@doi [ApJ] {10.3847/1538-4357/ac1869}, \href
  {https://ui.adsabs.harvard.edu/abs/2021ApJ...921...32J} {921, 32}

\bibitem[\protect\citeauthoryear{{Kalberla} \& {Haud}}{{Kalberla} \&
  {Haud}}{2006}]{Kalberla2006}
{Kalberla} P.~M.~W.,  {Haud} U.,  2006, \mn@doi [Astronomy \& Astrophysics]
  {10.1051/0004-6361:20054750}, \href
  {https://ui.adsabs.harvard.edu/abs/2006A&A...455..481K} {455, 481}

\bibitem[\protect\citeauthoryear{{Kallivayalil}, {van der Marel}, {Besla},
  {Anderson}  \& {Alcock}}{{Kallivayalil} et~al.}{2013}]{Kallivayalil2013}
{Kallivayalil} N.,  {van der Marel} R.~P.,  {Besla} G.,  {Anderson} J.,
  {Alcock} C.,  2013, \mn@doi [The Astrophysical Journal]
  {10.1088/0004-637X/764/2/161}, \href
  {https://ui.adsabs.harvard.edu/abs/2013ApJ...764..161K} {764, 161}

\bibitem[\protect\citeauthoryear{{Leisman} et~al.,}{{Leisman}
  et~al.}{2021}]{Leisman2021}
{Leisman} L.,  et~al., 2021, \mn@doi [AJ] {10.3847/1538-3881/ac2a38}, \href
  {https://ui.adsabs.harvard.edu/abs/2021AJ....162..274L} {162, 274}

\bibitem[\protect\citeauthoryear{{Lelli}, {McGaugh}  \& {Schombert}}{{Lelli}
  et~al.}{2016}]{Lelli2016}
{Lelli} F.,  {McGaugh} S.~S.,   {Schombert} J.~M.,  2016, \mn@doi [\aj]
  {10.3847/0004-6256/152/6/157}, \href
  {https://ui.adsabs.harvard.edu/abs/2016AJ....152..157L} {152, 157}

\bibitem[\protect\citeauthoryear{{Li}, {Hammer}, {Babusiaux}, {Pawlowski},
  {Yang}, {Arenou}, {Du}  \& {Wang}}{{Li} et~al.}{2021}]{Li2021}
{Li} H.,  {Hammer} F.,  {Babusiaux} C.,  {Pawlowski} M.~S.,  {Yang} Y.,
  {Arenou} F.,  {Du} C.,   {Wang} J.,  2021, \mn@doi [ApJ]
  {10.3847/1538-4357/ac0436}, \href
  {https://ui.adsabs.harvard.edu/abs/2021ApJ...916....8L} {916, 8}

\bibitem[\protect\citeauthoryear{{Longeard} et~al.,}{{Longeard}
  et~al.}{2022}]{Longeard2022}
{Longeard} N.,  et~al., 2022, \mn@doi [MNRAS] {10.1093/mnras/stac1827}, \href
  {https://ui.adsabs.harvard.edu/abs/2022MNRAS.516.2348L} {516, 2348}

\bibitem[\protect\citeauthoryear{{Mart{\'\i}nez-Garc{\'\i}a}, {del Pino}  \&
  {Aparicio}}{{Mart{\'\i}nez-Garc{\'\i}a} et~al.}{2023}]{Martinez-Garcia2023}
{Mart{\'\i}nez-Garc{\'\i}a} A.~M.,  {del Pino} A.,   {Aparicio} A.,  2023,
  \mn@doi [MNRAS] {10.1093/mnras/stac3305}, \href
  {https://ui.adsabs.harvard.edu/abs/2023MNRAS.518.3083M} {518, 3083}

\bibitem[\protect\citeauthoryear{{Mayer}}{{Mayer}}{2010}]{Mayer2010}
{Mayer} L.,  2010, \mn@doi [Advances in Astronomy] {10.1155/2010/278434}, \href
  {https://ui.adsabs.harvard.edu/abs/2010AdAst2010E..25M} {2010, 278434}

\bibitem[\protect\citeauthoryear{{Mayer}, {Mastropietro}, {Wadsley}, {Stadel}
  \& {Moore}}{{Mayer} et~al.}{2006}]{Mayer2006}
{Mayer} L.,  {Mastropietro} C.,  {Wadsley} J.,  {Stadel} J.,   {Moore} B.,
  2006, \mn@doi [\mnras] {10.1111/j.1365-2966.2006.10403.x}, \href
  {https://ui.adsabs.harvard.edu/abs/2006MNRAS.369.1021M} {369, 1021}

\bibitem[\protect\citeauthoryear{{McConnachie}}{{McConnachie}}{2012}]{McConnachie2012}
{McConnachie} A.~W.,  2012, \mn@doi [AJ] {10.1088/0004-6256/144/1/4}, \href
  {https://ui.adsabs.harvard.edu/abs/2012AJ....144....4M} {144, 4}

\bibitem[\protect\citeauthoryear{{McConnachie} \& {Venn}}{{McConnachie} \&
  {Venn}}{2020}]{McConnachie2020}
{McConnachie} A.~W.,  {Venn} K.~A.,  2020, \mn@doi [AJ]
  {10.3847/1538-3881/aba4ab}, \href
  {https://ui.adsabs.harvard.edu/abs/2020AJ....160..124M} {160, 124}

\bibitem[\protect\citeauthoryear{{Mu{\~n}oz}, {C{\^o}t{\'e}}, {Santana},
  {Geha}, {Simon}, {Oyarz{\'u}n}, {Stetson}  \& {Djorgovski}}{{Mu{\~n}oz}
  et~al.}{2018}]{Munoz2018}
{Mu{\~n}oz} R.~R.,  {C{\^o}t{\'e}} P.,  {Santana} F.~A.,  {Geha} M.,  {Simon}
  J.~D.,  {Oyarz{\'u}n} G.~A.,  {Stetson} P.~B.,   {Djorgovski} S.~G.,  2018,
  \mn@doi [ApJ] {10.3847/1538-4357/aac16b}, \href
  {https://ui.adsabs.harvard.edu/abs/2018ApJ...860...66M} {860, 66}

\bibitem[\protect\citeauthoryear{{Navarro}, {Frenk}  \& {White}}{{Navarro}
  et~al.}{1997}]{Navarro1997}
{Navarro} J.~F.,  {Frenk} C.~S.,   {White} S. D.~M.,  1997, \mn@doi [\apj]
  {10.1086/304888}, \href
  {https://ui.adsabs.harvard.edu/abs/1997ApJ...490..493N} {490, 493}

\bibitem[\protect\citeauthoryear{{Nidever}}{{Nidever}}{2014}]{Nidever2014}
{Nidever} D.~L.,  2014, in {Seigar} M.~S.,  {Treuthardt} P.,  eds,
  Astronomical Society of the Pacific Conference Series Vol. 480, Structure and
  Dynamics of Disk Galaxies. p.~27 (\mn@eprint {arXiv} {1310.6742})

\bibitem[\protect\citeauthoryear{{Pace}, {Erkal}  \& {Li}}{{Pace}
  et~al.}{2022}]{Pace2022}
{Pace} A.~B.,  {Erkal} D.,   {Li} T.~S.,  2022, \mn@doi [\apj]
  {10.3847/1538-4357/ac997b}, \href
  {https://ui.adsabs.harvard.edu/abs/2022ApJ...940..136P} {940, 136}

\bibitem[\protect\citeauthoryear{{Pawlowski}, {Pflamm-Altenburg}  \&
  {Kroupa}}{{Pawlowski} et~al.}{2012}]{Pawlowski2012}
{Pawlowski} M.~S.,  {Pflamm-Altenburg} J.,   {Kroupa} P.,  2012, \mn@doi
  [\mnras] {10.1111/j.1365-2966.2012.20937.x}, \href
  {https://ui.adsabs.harvard.edu/abs/2012MNRAS.423.1109P} {423, 1109}

\bibitem[\protect\citeauthoryear{{Putman}, {Zheng}, {Price-Whelan}, {Grcevich},
  {Johnson}, {Tollerud}  \& {Peek}}{{Putman} et~al.}{2021}]{Putman2021}
{Putman} M.~E.,  {Zheng} Y.,  {Price-Whelan} A.~M.,  {Grcevich} J.,  {Johnson}
  A.~C.,  {Tollerud} E.,   {Peek} J. E.~G.,  2021, \mn@doi [\apj]
  {10.3847/1538-4357/abe391}, \href
  {https://ui.adsabs.harvard.edu/abs/2021ApJ...913...53P} {913, 53}

\bibitem[\protect\citeauthoryear{{Read}, {Walker}  \& {Steger}}{{Read}
  et~al.}{2019}]{Read2019}
{Read} J.~I.,  {Walker} M.~G.,   {Steger} P.,  2019, \mn@doi [MNRAS]
  {10.1093/mnras/sty3404}, \href
  {https://ui.adsabs.harvard.edu/abs/2019MNRAS.484.1401R} {484, 1401}

\bibitem[\protect\citeauthoryear{{Rocha}, {Peter}  \& {Bullock}}{{Rocha}
  et~al.}{2012}]{Rocha2012}
{Rocha} M.,  {Peter} A. H.~G.,   {Bullock} J.,  2012, \mn@doi [MNRAS]
  {10.1111/j.1365-2966.2012.21432.x}, \href
  {https://ui.adsabs.harvard.edu/abs/2012MNRAS.425..231R} {425, 231}

\bibitem[\protect\citeauthoryear{{Roederer}, {Pace}, {Placco}, {Caldwell},
  {Koposov}, {Mateo}, {Olszewski}  \& {Walker}}{{Roederer}
  et~al.}{2023}]{Roederer2023}
{Roederer} I.~U.,  {Pace} A.~B.,  {Placco} V.~M.,  {Caldwell} N.,  {Koposov}
  S.~E.,  {Mateo} M.,  {Olszewski} E.~W.,   {Walker} M.~G.,  2023, arXiv
  e-prints, \href {https://ui.adsabs.harvard.edu/abs/2023arXiv230702585R} {p.
  arXiv:2307.02585}

\bibitem[\protect\citeauthoryear{{Salem}, {Besla}, {Bryan}, {Putman}, {van der
  Marel}  \& {Tonnesen}}{{Salem} et~al.}{2015}]{Salem2015}
{Salem} M.,  {Besla} G.,  {Bryan} G.,  {Putman} M.,  {van der Marel} R.~P.,
  {Tonnesen} S.,  2015, \mn@doi [ApJ] {10.1088/0004-637X/815/1/77}, \href
  {http://adsabs.harvard.edu/abs/2015ApJ...815...77S} {815, 77}

\bibitem[\protect\citeauthoryear{{Sestito} et~al.,}{{Sestito}
  et~al.}{2023}]{Sestito2023}
{Sestito} F.,  et~al., 2023, \mn@doi [arXiv e-prints]
  {10.48550/arXiv.2301.13214}, \href
  {https://ui.adsabs.harvard.edu/abs/2023arXiv230113214S} {p. arXiv:2301.13214}

\bibitem[\protect\citeauthoryear{{Springel}}{{Springel}}{2005}]{Springel2005}
{Springel} V.,  2005, \mn@doi [\mnras] {10.1111/j.1365-2966.2005.09655.x},
  \href {https://ui.adsabs.harvard.edu/abs/2005MNRAS.364.1105S} {364, 1105}

\bibitem[\protect\citeauthoryear{{Taibi}, {Pawlowski}, {Khoperskov},
  {Steinmetz}  \& {Libeskind}}{{Taibi} et~al.}{2023}]{Taibi2023}
{Taibi} S.,  {Pawlowski} M.~S.,  {Khoperskov} S.,  {Steinmetz} M.,
  {Libeskind} N.~I.,  2023, \mn@doi [arXiv e-prints]
  {10.48550/arXiv.2310.13521}, \href
  {https://ui.adsabs.harvard.edu/abs/2023arXiv231013521T} {p. arXiv:2310.13521}

\bibitem[\protect\citeauthoryear{{Tepper-Garc{\'\i}a}, {Bland-Hawthorn},
  {Pawlowski}  \& {Fritz}}{{Tepper-Garc{\'\i}a}
  et~al.}{2019}]{Tepper-garcia2019}
{Tepper-Garc{\'\i}a} T.,  {Bland-Hawthorn} J.,  {Pawlowski} M.~S.,   {Fritz}
  T.~K.,  2019, \mn@doi [MNRAS] {10.1093/mnras/stz1659}, \href
  {https://ui.adsabs.harvard.edu/abs/2019MNRAS.488..918T} {488, 918}

\bibitem[\protect\citeauthoryear{{Torrealba}, {Koposov}, {Belokurov}  \&
  {Irwin}}{{Torrealba} et~al.}{2016}]{Torrealba2016}
{Torrealba} G.,  {Koposov} S.~E.,  {Belokurov} V.,   {Irwin} M.,  2016, \mn@doi
  [\mnras] {10.1093/mnras/stw733}, \href
  {https://ui.adsabs.harvard.edu/abs/2016MNRAS.459.2370T} {459, 2370}

\bibitem[\protect\citeauthoryear{{Torrealba} et~al.,}{{Torrealba}
  et~al.}{2019}]{Torrealba2019}
{Torrealba} G.,  et~al., 2019, \mn@doi [MNRAS] {10.1093/mnras/stz1624}, \href
  {https://ui.adsabs.harvard.edu/abs/2019MNRAS.488.2743T} {488, 2743}

\bibitem[\protect\citeauthoryear{{Vasiliev}}{{Vasiliev}}{2013}]{Vasiliev2013}
{Vasiliev} E.,  2013, \mn@doi [MNRAS] {10.1093/mnras/stt1235}, \href
  {http://adsabs.harvard.edu/abs/2013MNRAS.434.3174V} {434, 3174}

\bibitem[\protect\citeauthoryear{{Vasiliev}}{{Vasiliev}}{2019}]{Vasiliev2019}
{Vasiliev} E.,  2019, \mn@doi [MNRAS] {10.1093/mnras/sty2672}, \href
  {https://ui.adsabs.harvard.edu/abs/2019MNRAS.482.1525V} {482, 1525}

\bibitem[\protect\citeauthoryear{{Walker}, {Mateo}, {Olszewski}, {Gnedin},
  {Wang}, {Sen}  \& {Woodroofe}}{{Walker} et~al.}{2007}]{Walker2007}
{Walker} M.~G.,  {Mateo} M.,  {Olszewski} E.~W.,  {Gnedin} O.~Y.,  {Wang} X.,
  {Sen} B.,   {Woodroofe} M.,  2007, \mn@doi [ApJL] {10.1086/521998}, \href
  {https://ui.adsabs.harvard.edu/abs/2007ApJ...667L..53W} {667, L53}

\bibitem[\protect\citeauthoryear{{Walker}, {Mateo}, {Olszewski},
  {Pe{\~n}arrubia}, {Evans}  \& {Gilmore}}{{Walker} et~al.}{2009}]{Walker2009}
{Walker} M.~G.,  {Mateo} M.,  {Olszewski} E.~W.,  {Pe{\~n}arrubia} J.,  {Evans}
  N.~W.,   {Gilmore} G.,  2009, \mn@doi [ApJ] {10.1088/0004-637X/704/2/1274},
  \href {https://ui.adsabs.harvard.edu/abs/2009ApJ...704.1274W} {704, 1274}

\bibitem[\protect\citeauthoryear{{Waller} et~al.,}{{Waller}
  et~al.}{2023}]{Waller2023}
{Waller} F.,  et~al., 2023, \mn@doi [MNRAS] {10.1093/mnras/stac3563}, \href
  {https://ui.adsabs.harvard.edu/abs/2023MNRAS.519.1349W} {519, 1349}

\bibitem[\protect\citeauthoryear{{Wang}, {Hammer}, {Yang}, {Ripepi}, {Cioni},
  {Puech}  \& {Flores}}{{Wang} et~al.}{2019}]{Wang2019}
{Wang} J.,  {Hammer} F.,  {Yang} Y.,  {Ripepi} V.,  {Cioni} M.-R.~L.,  {Puech}
  M.,   {Flores} H.,  2019, \mn@doi [MNRAS] {10.1093/mnras/stz1274}, \href
  {https://ui.adsabs.harvard.edu/abs/2019MNRAS.486.5907W} {486, 5907}

\bibitem[\protect\citeauthoryear{{Wang}, {Hammer}  \& {Yang}}{{Wang}
  et~al.}{2022}]{Wang2022}
{Wang} J.,  {Hammer} F.,   {Yang} Y.,  2022, \mn@doi [MNRAS]
  {10.1093/mnras/stac1640}, \href
  {https://ui.adsabs.harvard.edu/abs/2022MNRAS.515..940W} {515, 940}

\bibitem[\protect\citeauthoryear{{Westfall}, {Majewski}, {Ostheimer},
  {Frinchaboy}, {Kunkel}, {Patterson}  \& {Link}}{{Westfall}
  et~al.}{2006}]{Westfall2006}
{Westfall} K.~B.,  {Majewski} S.~R.,  {Ostheimer} J.~C.,  {Frinchaboy} P.~M.,
  {Kunkel} W.~E.,  {Patterson} R.~J.,   {Link} R.,  2006, \mn@doi [The
  Astronomical Journal] {10.1086/496975}, \href
  {https://ui.adsabs.harvard.edu/abs/2006AJ....131..375W} {131, 375}

\bibitem[\protect\citeauthoryear{{Wolf}, {Martinez}, {Bullock}, {Kaplinghat},
  {Geha}, {Mu{\~n}oz}, {Simon}  \& {Avedo}}{{Wolf} et~al.}{2010}]{Wolf2010}
{Wolf} J.,  {Martinez} G.~D.,  {Bullock} J.~S.,  {Kaplinghat} M.,  {Geha} M.,
  {Mu{\~n}oz} R.~R.,  {Simon} J.~D.,   {Avedo} F.~F.,  2010, MNRAS, \href
  {http://cdsads.u-strasbg.fr/abs/2010MNRAS.406.1220W} {406, 1220}

\bibitem[\protect\citeauthoryear{{Yang}, {Hammer}, {Fouquet}, {Flores},
  {Puech}, {Pawlowski}  \& {Kroupa}}{{Yang} et~al.}{2014}]{Yang2014}
{Yang} Y.,  {Hammer} F.,  {Fouquet} S.,  {Flores} H.,  {Puech} M.,  {Pawlowski}
  M.~S.,   {Kroupa} P.,  2014, \mn@doi [MNRAS] {10.1093/mnras/stu931}, \href
  {http://cdsads.u-strasbg.fr/abs/2014MNRAS.442.2419Y} {442, 2419}

\bibitem[\protect\citeauthoryear{{Yang}, {Ianjamasimanana}, {Hammer}, {Higgs},
  {Namumba}, {Carignan}, {J{\'o}zsa}  \& {McConnachie}}{{Yang}
  et~al.}{2022a}]{Yang2022b}
{Yang} Y.,  {Ianjamasimanana} R.,  {Hammer} F.,  {Higgs} C.,  {Namumba} B.,
  {Carignan} C.,  {J{\'o}zsa} G. I.~G.,   {McConnachie} A.~W.,  2022a, arXiv
  e-prints, \href {https://ui.adsabs.harvard.edu/abs/2022arXiv220403662Y} {p.
  arXiv:2204.03662}

\bibitem[\protect\citeauthoryear{{Yang}, {Hammer}, {Jiao}  \&
  {Pawlowski}}{{Yang} et~al.}{2022b}]{Yang2022a}
{Yang} Y.,  {Hammer} F.,  {Jiao} Y.,   {Pawlowski} M.~S.,  2022b, \mn@doi
  [MNRAS] {10.1093/mnras/stac644}, \href
  {https://ui.adsabs.harvard.edu/abs/2022MNRAS.512.4171Y} {512, 4171}

\bibitem[\protect\citeauthoryear{{de Boer} et~al.,}{{de Boer}
  et~al.}{2012}]{deBoer2012}
{de Boer} T.~J.~L.,  et~al., 2012, \mn@doi [\aap]
  {10.1051/0004-6361/201118378}, \href
  {https://ui.adsabs.harvard.edu/abs/2012A&A...539A.103D} {539, A103}

\makeatother
\end{thebibliography}

\begin{appendix}

\section{Stability of initial conditions for the dwarf progenitors}
\label{sec:ICs}

\begin{figure*}
\centering
\includegraphics[width=17cm]{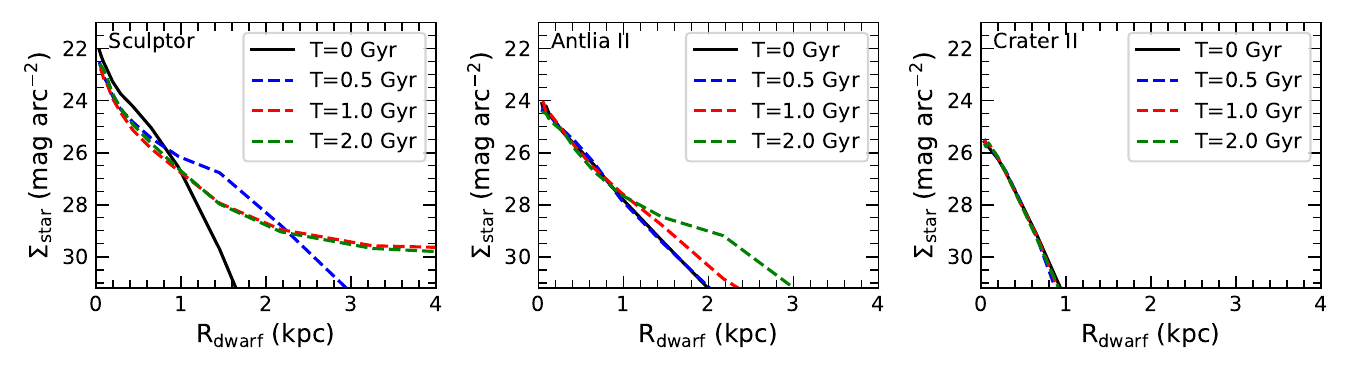}
\includegraphics[width=17cm]{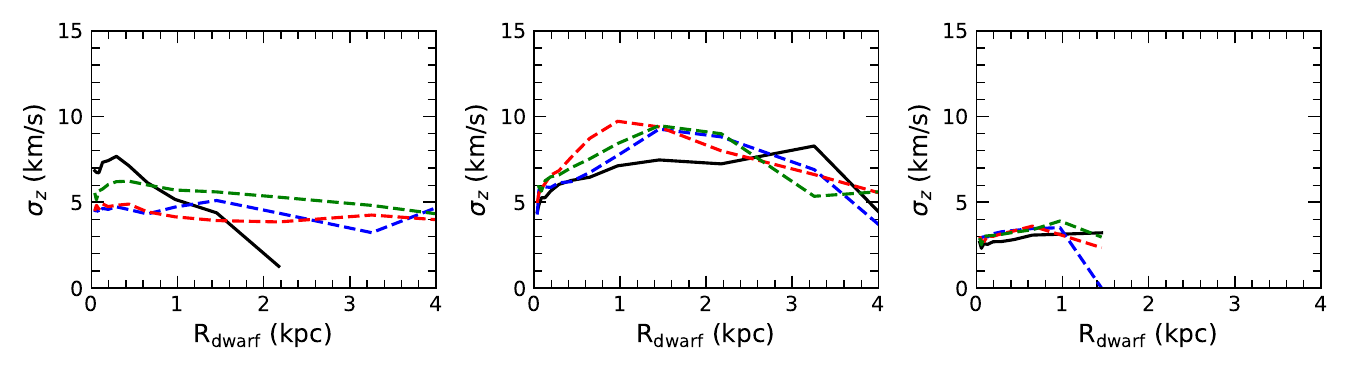}
\caption{Evolution of the surface brightness and projected velocity 
dispersion profiles for the dwarf progenitors assumed to be in isolation.}
\label{ICs}
\end{figure*}

Figure~\ref{ICs} shows the evolution of the projected surface brightness and line-of-sight velocity dispersion for the progenitors of the three dwarfs assumed to be in isolation for 2 Gyr. In order to
calculate the surface brightness, we have assumed their distance to be 517 kpc (the starting position of simulation models). The high surface mass density of gas and high feedback strength used in Sculptor lead to relatively large variations in surface brightness at large radii. This variation is caused by the strong adopted feedback and star formation, which result in large variations at the outskirts, i.e., beyond 4 times the half-light radius. The low gas mass for Crater II and median feedback strength for Antlia II lead to  relatively stable surface brightness and velocity dispersion profile. However, for all gas-rich dwarf progenitors the velocity dispersion does not change by more than 2-3 $\rm km s^{-1}$, and it always shows a relatively flat profile. The effect of gas perturbation onto stars are much more effective when the dwarfs fall into  the Galactic corona. This is because hydrodynamical processes such as ram-pressure, tidal shocks, and turbulence, make the stellar component much more perturbed by turbulence during the time the dwarf progenitors are dominated by the gas interacting with the Galactic coronae.


\section{Resolution dependency of parameters to reproduce dwarf properties}
\label{sec:degenerate} 

Ideally, it would be better to use the same resolution for gas particles of the Galactic corona and of the dwarf progenitors. But this is too expensive in calculation time because the dwarfs are too tiny when compared to the Milky Way. It would require 10, 15 and 30 times more particles for Milky Way model than what has been used in our current modeling with $\sim 30$ million particles. Simulations with such high numbers of particles are not feasible currently. Therefore, we have used lower resolution modeling in our current work to investigate the parameter space, and to search for the best matched dwarf modeling.
However, we have tested the impact of resolution by increasing by a factor 3 the number of particles of the Milky Way halo gas. We have found that it leads to larger elapsed times (by about one Gyr) for dwarf progenitors to be fully gas-stripped.

\begin{figure}
\centering
\includegraphics[width=8cm]{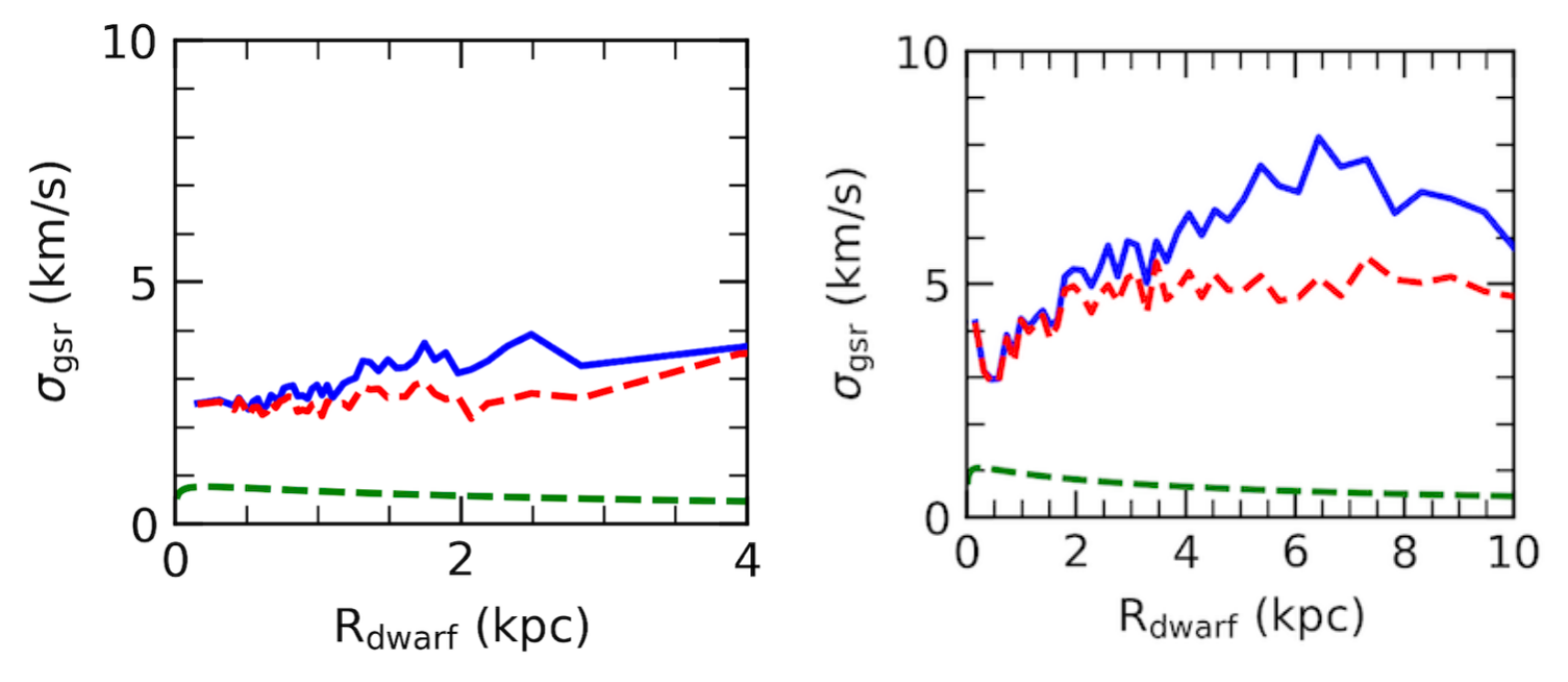}
\caption{Velocity dispersion profiles for Crater II (left panel) and Antlia II (right panel) with a higher resolution for the MW gas particles. Increasing resolution of Galactic model leads to 
delaying the gas stripping. By tuning other parameters, we can strip the gas of dwarf  at slightly different observation times.
For the Crater II analog, the time delay does not change the velocity dispersion too much. For Antlia II, decreasing the gas fraction by a factor of 1.5 shortens the gas stripping time, allowing it to occur near pericenter. The velocity dispersion is roughly recovered as it is shown in the right panel.} 
\label{ResMW}
\end{figure}

Figure~\ref{ResMW} shows how one may need to reinvestigate the parameter space for matching observations after performing higher resolution simulations. This illustrates the interplay between these 
parameters and the resolution of the MW gas particles in the simulation.

For Crater II, increasing the MW particle resolution by a factor 3 results in gas removal at $\sim 4.35$ Gyr (and a distance of 102 kpc), compared to $\sim 3.4$ Gyr in the fiducial simulation.
However, the velocity dispersion profile is similar to the low resolution one (see left panel of Figure~\ref{ResMW}). For the Antlia II simulation, increasing the MW particle resolution also delays the gas stripping. We find that by decreasing the initial gas fraction by a factor of 1.5, the gas can be stripped near pericenter, and the observed velocity dispersion is rather well recovered (see right panel of Figure~\ref{ResMW}) as it was in the low resolution simulation (see Figure~\ref{AntII}). 
Nevertheless, Figure~\ref{ResMW} shows that one may recover the observed properties of dwarfs at higher numerical resolutions.

\end{appendix}

\end{document}